\documentclass[preprint,12pt]{elsarticle}

\usepackage{amssymb}

\usepackage{color}

\journal{Physica A}

\begin{document}

\begin{frontmatter}



\title{Intermittent gravity-driven flow of grains through narrow pipes \tnoteref{label_note_copyright} \tnoteref{label_note_doi}}

\tnotetext[label_note_copyright]{\copyright 2016. This manuscript version is made available under the CC-BY-NC-ND 4.0 license http://creativecommons.org/licenses/by-nc-nd/4.0/}

\tnotetext[label_note_doi]{Accepted Manuscript for Physica A, v. 465, p. 725-741, DOI:10.1016/j.physa.2016.08.071}


\author[label_carlos]{Carlos A. Alvarez}
\author[label_erick]{Erick de Moraes Franklin}

\address[label_erick]{Faculty of Mechanical Engineering - University of Campinas - UNICAMP\\
e-mail: franklin@fem.unicamp.br\\
Rua Mendeleyev, 200 - Campinas - SP - CEP: 13083-970\\
Brazil}

\address[label_carlos]{Faculty of Mechanical Engineering - University of Campinas - UNICAMP\\
e-mail: calvarez@fem.unicamp.br\\
Rua Mendeleyev, 200 - Campinas - SP - CEP: 13083-970\\
Brazil}

\begin{abstract}
\begin{sloppypar}
Grain flows through pipes are frequently found in various settings, such as in pharmaceutical, chemical, petroleum, mining and food industries. In the case of size-constrained gravitational flows, density waves consisting of alternating high- and low-compactness regions may appear. This study investigates experimentally the dynamics of density waves that appear in gravitational flows of fine grains through vertical and slightly inclined pipes. The experimental device consisted of a transparent glass pipe through which different populations of glass spheres flowed driven by gravity. Our experiments were performed under controlled ambient temperature and relative humidity, and the granular flow was filmed with a high-speed camera. Experimental results concerning the length scales and celerities of density waves are presented, together with a one-dimensional model and a linear stability analysis. The analysis exhibits the presence of a long-wavelength instability, with the most unstable mode and a cut-off wavenumber whose values are in agreement with the experimental results.
\end{sloppypar}
\end{abstract}


\end{frontmatter}



\section{Introduction}

Granular materials play an important role in our daily lives, for instance, arid regions occupy about $20 \%$ of Earth's surface, the global annual production of grains and aggregates is approximately ten billion metric tons, and the processing of granular media consumes roughly $10 \%$ of all the energy produced worldwide \cite{Duran}. As a result, gravitational flows of these materials are frequently observed in nature and industry. However, the behavior of granular flows is not well understood, as granular matter is a discrete medium whose rheology is unknown. Given the importance of granular flows, considerable work has been done to understand their dynamics and instabilities \cite{Campbell,Elbelrhiti,Franklin_6,GDR_midi,Jaeger}.

Gravitational grain flows in pipes are common in industry. Some examples are the transport of grains in the food industry, the transport of sand in civil constructions, and the transport of powders in the chemical and pharmaceutical industries. When the grains and the tube diameter are size-constrained, granular flow may give rise to instabilities. These instabilities consist of alternating high- and low-compactness regions (regions of high and low grain concentration, respectively), and are characterized by intermittency, oscillating patterns and even blockages \cite{Aider,Bertho_1,Raafat}. Although this instability may appear under vacuum conditions \cite{Savage,Wang}, in the case of fine grains these patterns are recognized as the result of the interaction between small-size falling grains and trapped air.

Lee \cite{Lee} investigated the density waves in granular flows through vertical tubes and hoppers using analytical techniques and numerical simulations. The author used mass and momentum equations to describe density and velocity fields. In the equations, the law of friction proposed by Bagnold \cite{Bagnold_4} was employed and the effects of both air pressure and drag caused on grains were neglected. For the vertical tubes, the author found that kinetic waves exist and partially obtained a dispersion relation for the dynamic waves, which he did not solve. The numerical simulations were performed using molecular dynamics (MD), and the author found indications that the density waves are of kinetic nature. However, because air effects (pressure and drag) were absent in both the stability analysis and the numerical simulation, the results are not suitable in the case of fine grains in narrow pipes. In addition, in the case of density waves, the grains in high-density regions are in permanent contact; therefore, MD is not an adequate method since it assumes binary and instantaneous contacts.

Raafat et al. \cite{Raafat} studied the formation of density waves in pipes experimentally. The experiments were performed in a $1.3\,m$ long tube with an internal diameter $D$ of $2.9\,mm$ using glass splinters and glass beads with mean grain diameter $d$ of $0.09\,mm$ to $0.2\,mm$ and $0.2\,mm$, respectively. They observed density waves for moderate grain flow rate and when the ratio between the pipe and the grain diameter is $6 \leq D/d \leq 30$. Furthermore, they proposed that the friction between the grains and the forces between the trapped air and the grains are responsible for the density waves.

Aider et al. \cite{Aider} presented an experimental study of the granular flow patterns in vertical pipes. The experiments were performed in a tube similar to that of Raafat et al. \cite{Raafat} using glass beads with mean diameter of $125\, \mu m$. The density variations were measured using a linear CCD (charge coupled device) camera with frequencies of up to $2\,kHz$. Aider et al. \cite{Aider} observed that the density waves consisted of high-compactness plugs ($c\approx 60\%$, where $c$ is the compactness) separated by low-density regions; furthermore, the density waves appeared when the grain flow rate $\dot{m}$ was $1.5\,g/s\,-\, 2.5\,g/s$ (oscillating waves) or $2.5\,g/s\,-\,5\,g/s$ (propagative waves). The authors also noted that humidity $H$ must be within $35\%$ and $75\%$, otherwise the grains clogged the tube due to capillary forces ($H>75\%$) or due to electrostatic forces ($H<35\%$).

Bertho et al. \cite{Bertho_1} presented experiments on density waves using an experimental set-up similar to that of Raafat et al. \cite{Raafat} and Aider et al. \cite{Aider}. The vertical tube ($D=3\,mm$, $1.25 \,m$ long) and the glass beads ($d\,=\,125\, \mu m$ glass beads) were more or less the same as those of Aider et al. \cite{Aider}, and a linear CCD camera was used. In addition, capacitance sensors were used to measure the compactness of grains at two different locations, and the pressure distribution was also measured. The experimental data showed that the characteristic length of the high-compactness regions of the density wave regime is in the order of $10\,mm$.

Recently, Franklin and Alvarez \cite{Franklin_7} presented a linear stability analysis and experimental results for the vertical chute of grains in a narrow pipe. They found a dimensional dispersion relation to be solved numerically, and the analysis was limited to some small ranges of grains and pipes. The experiments were performed in a $1\,m$ long glass tube of $3\,mm$ internal diameter aligned vertically, and the grains consisted of glass beads of specific mass $\rho_s\,=\,2500\,kg/m^3$ divided in two different populations: grains with diameter within $212 \,\mu m\,\leq \, d \,\leq \, 300 \,\mu m$ and within $106 \,\mu m\,\leq \, d \,\leq \, 212 \,\mu m$. Franklin and Alvarez \cite{Franklin_7} reported the existence of granular plugs with length in the range $3\, <\,\lambda /D\, <\, 11$, where $\lambda$ is the plug length.

Numerical studies on intermittent granular flows in pipes have been carried out in recent years. Ellingsen et al. \cite{Ellingsen} studied the gravitational flow of grains through a narrow pipe under vacuum conditions. They performed numerical simulations based on a one-dimensional model for the granular flow where the collisions were modeled using two coefficients of restitution, one among grains and the other between the grains and the pipe walls. A narrow pipe was assumed and periodic boundary conditions were employed. The numerical results showed that granular waves could form in the absence of air if the dissipation caused by the collisions among the grains was smaller than that between the grains and the walls. However, the proposed model cannot predict the wavelength of the density waves in the presence of interstitial gas. Verb{\"u}cheln et al. \cite{Verbucheln}, using particle-based numerical simulations, found that density waves depend on the mass flow rate, particle distribution, and geometrical parameters of the pipe (pipe diameter and wall roughness). The authors reported plugs moving with constant velocity along the pipe. Moreover, they observed that the plugs do not break up with the impact of the smaller particle groups that fall onto it. In other words, the frictional forces that yield the arches leading to plug formation are strong enough to sustain the downward pressure on the granular column. The numerical simulations were conducted with different pipe diameters, but the granular plugs usually appeared when the diameter was 3 $mm$.

The density waves appearing in gravitational flows have similarities with the waves appearing in the vertical pneumatic conveying in dense regimes \cite{Konrad_1,Borzone,Konrad_2,Jaworski}. Konrad \cite{Konrad_1} presented a review on the pneumatic conveying of grains in horizontal and vertical tubes. For dense upward flows in vertical tubes, the author explained the formation of plugs as the interaction between the pressure differences between consecutive air bubbles (low compactness regions) and the weight of grains within the granular plug. The former is directed upwards, whereas the latter is subjected to the Janssen effect and results in a drag force directed downwards.

Borzone and Kinzing \cite {Borzone} studied the formation of plugs in a vertical pipe of $25.4\,mm$ internal diameter. Coal particles with diameters in the range $16\, \mu m \,\leq\,d\,\leq\, 63\, \mu m$ were pneumatically conveyed, and the pressure drop and the plug length were measured. The authors reported granular plugs with lengths in the range $2\, <\,\lambda /D\, <\, 15$.

Jaworski and Dyakowski \cite{Jaworski} studied the granular plugs in pneumatic conveying systems. The experiments were conducted in a $7\,m$ long horizontal section and in a $3\,m$ long vertical section, both of $57\,mm$ internal diameter, using polyamide chips measuring $3\,mm\,\times\, 3\,mm\,\times\, 1\,mm$. They measured the density waves with a high-speed camera and an Electrical Capacitance Tomography (ECT) device. In the vertical test section, Jaworski and Dyakowski  \cite{Jaworski} reported the existence of series of granular plugs, the length of each plug ranging between $2D$ and $4D$.

The objective of the present study is to determine the wavelengths and celerities of density waves that appear when fine grains fall through vertical and slightly inclined pipes. An experimental investigation was undertaken, using high-frequency movies to measure wavelengths and celerities. Our experiments were performed under controlled ambient temperature and relative humidity, and the granular flow was filmed with a high-speed camera. Moreover, this paper presents a one-dimensional flow model based on the work of Bertho et al. \cite{Bertho_2} with the inclusion of closure equations for the friction terms, and also a linear stability analysis. The flow model is made dimensionless and the stability analysis takes into consideration the main mechanisms involved, namely the Janssen effect, the interaction between the grains and the air, and gravity, and the results are then compared to the experimental data.

The next sections describe the experimental set-up and the experimental results. The following sections present the physics and the main equations of the one-dimensional model, the stability analysis of the granular flow, and the discussion of the main results. The conclusion section follows.

\section{Experimental device}

\begin{figure}[h!]
\begin{center}
\includegraphics[scale=0.12]{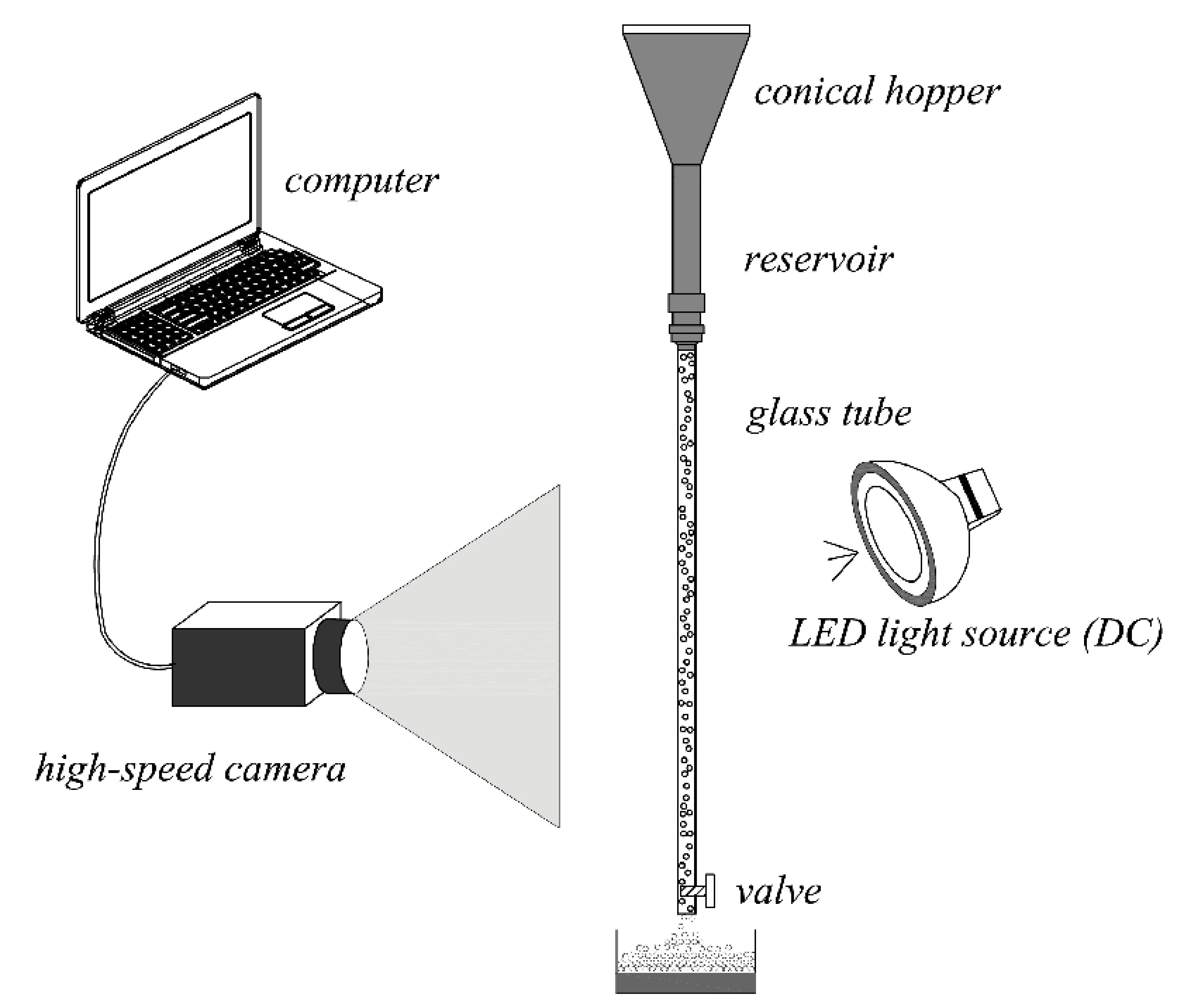}
\caption{Experimental set-up}
\label{fig:setup}
\end{center}
\end{figure}

The experimental device consisted of a conical hopper with an opening angle of 60$^o$, a reservoir, a 1.0 m long vertical glass pipe with an internal diameter $D$ = 3 mm (test section), and an exit valve (Fig. \ref{fig:setup}).

The grains consisted of glass beads of specific mass $\rho_s\,=\,2500\,kg/m^3$ divided in two different populations, namely: grain diameters within $106 \,\mu m\,\leq \, d \,\leq \, 212 \,\mu m$ and within $212 \,\mu m\,\leq \, d \,\leq \, 300 \,\mu m$. The tube was vertically aligned (within $\pm\,5^o$) and both the reservoir entrance and the exit valve were at atmospheric pressure. The temperature and the humidity measurements were within $\pm$ 0.5 $^{\circ}C$ and $\pm$ 2.5$\%$, respectively. These two parameters were controlled throughout every test. 

\begin{figure}[ht!]
\begin{minipage}[c]{0.5\textwidth}
\begin{tabular}{c}
\includegraphics[width=\linewidth]{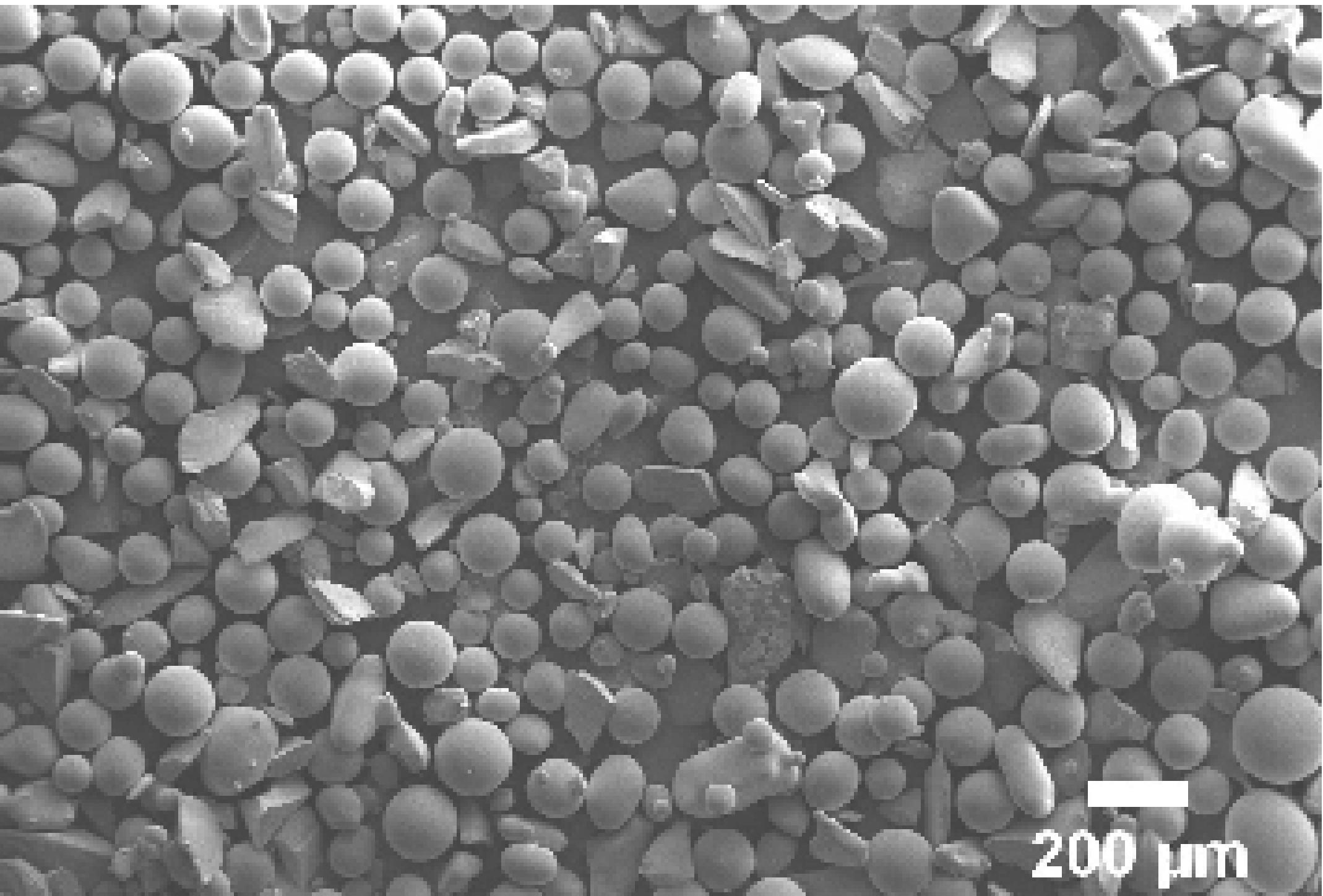}\\
      (a)
\end{tabular}
\end{minipage} \hfill
\begin{minipage}[c]{0.5\textwidth}
\hspace{\fill}
\begin{tabular}{c}
\includegraphics[width=\linewidth]{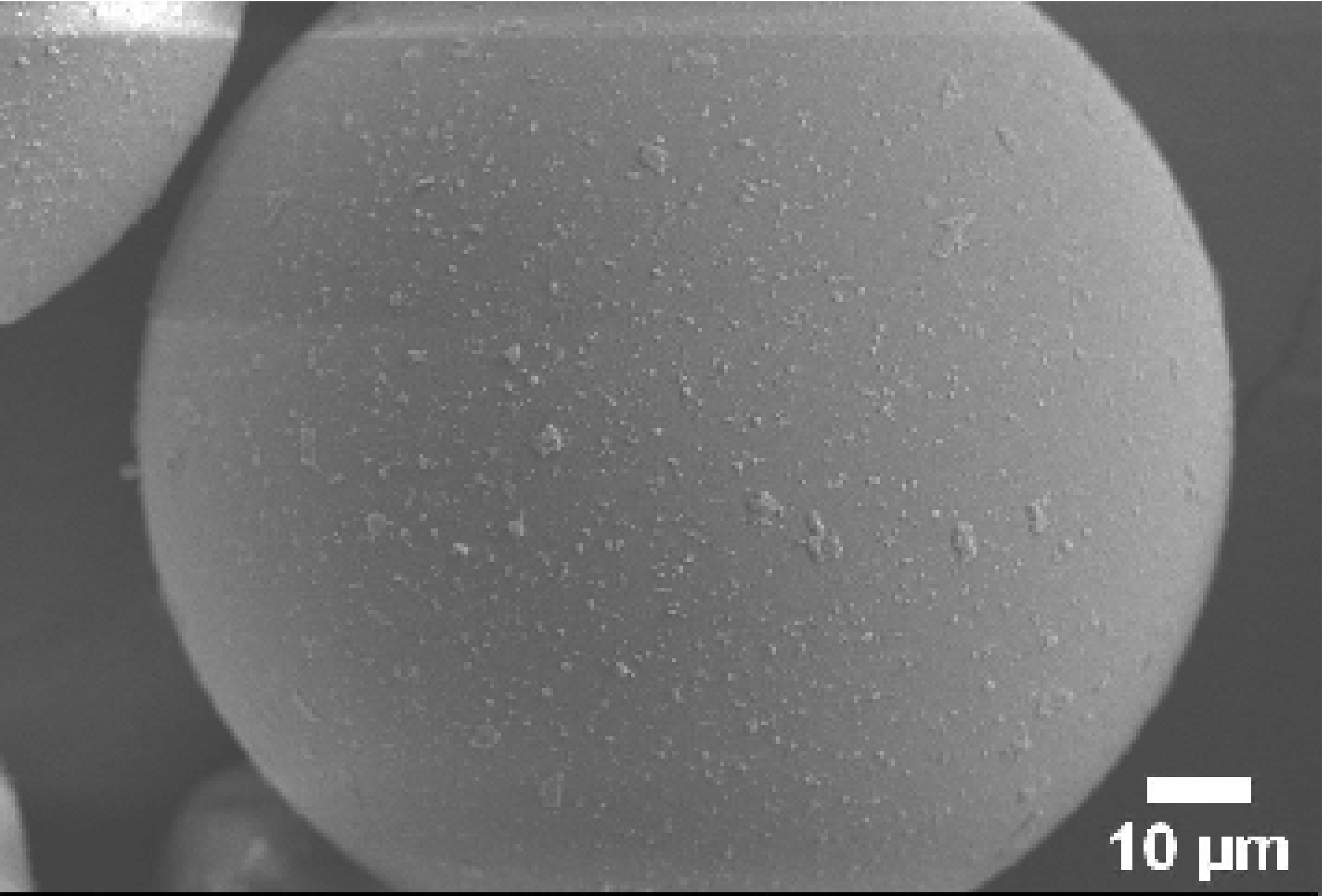}\\
      (b)
\end{tabular}
\end{minipage}
\begin{minipage}[c]{0.5\textwidth}

\begin{tabular}{c}
\includegraphics[width=\linewidth]{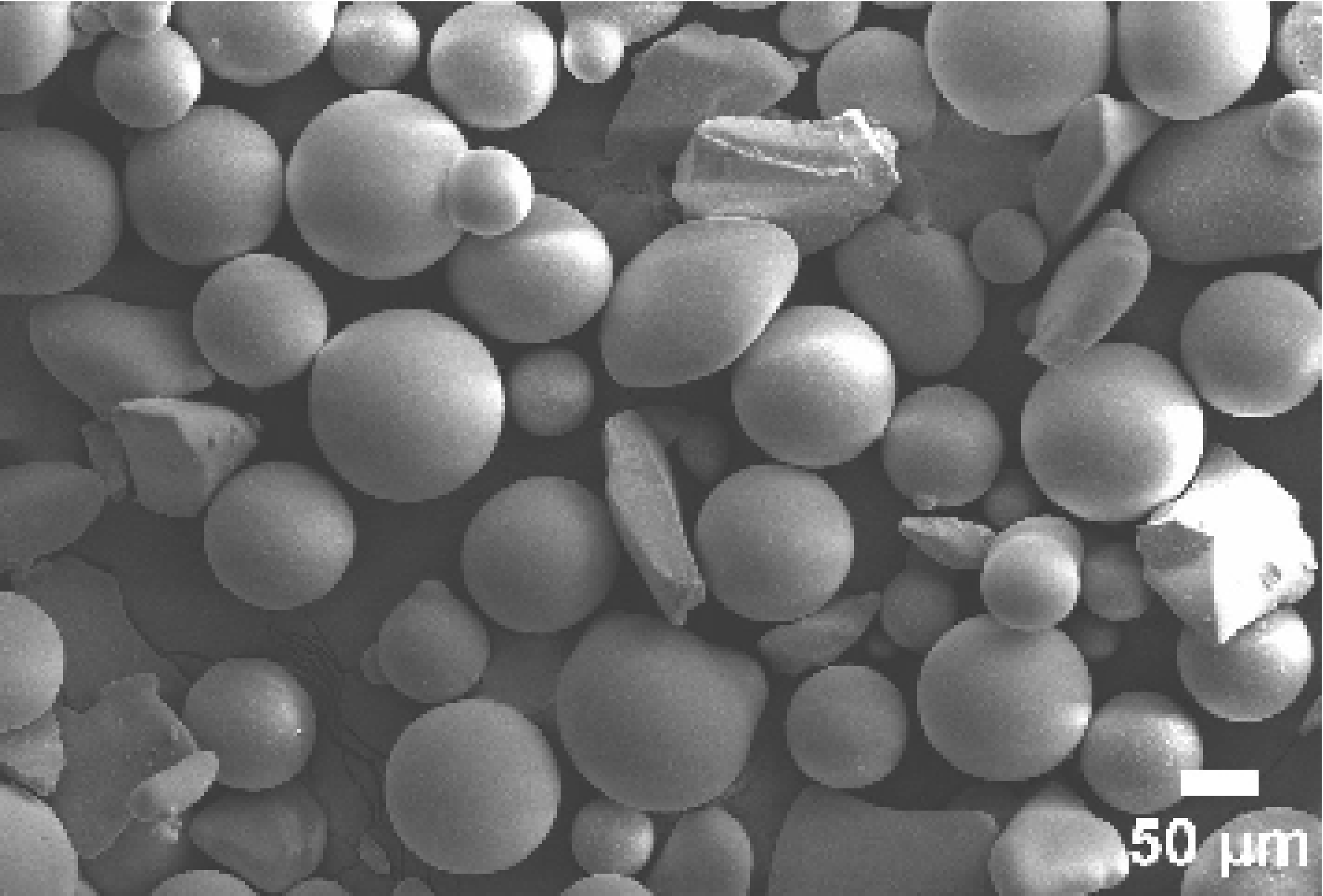}\\
      (c)
\end{tabular}
\end{minipage}
\begin{minipage}[c]{0.5\textwidth}
\hspace{\fill}
\begin{tabular}{c}
\includegraphics[width=\linewidth]{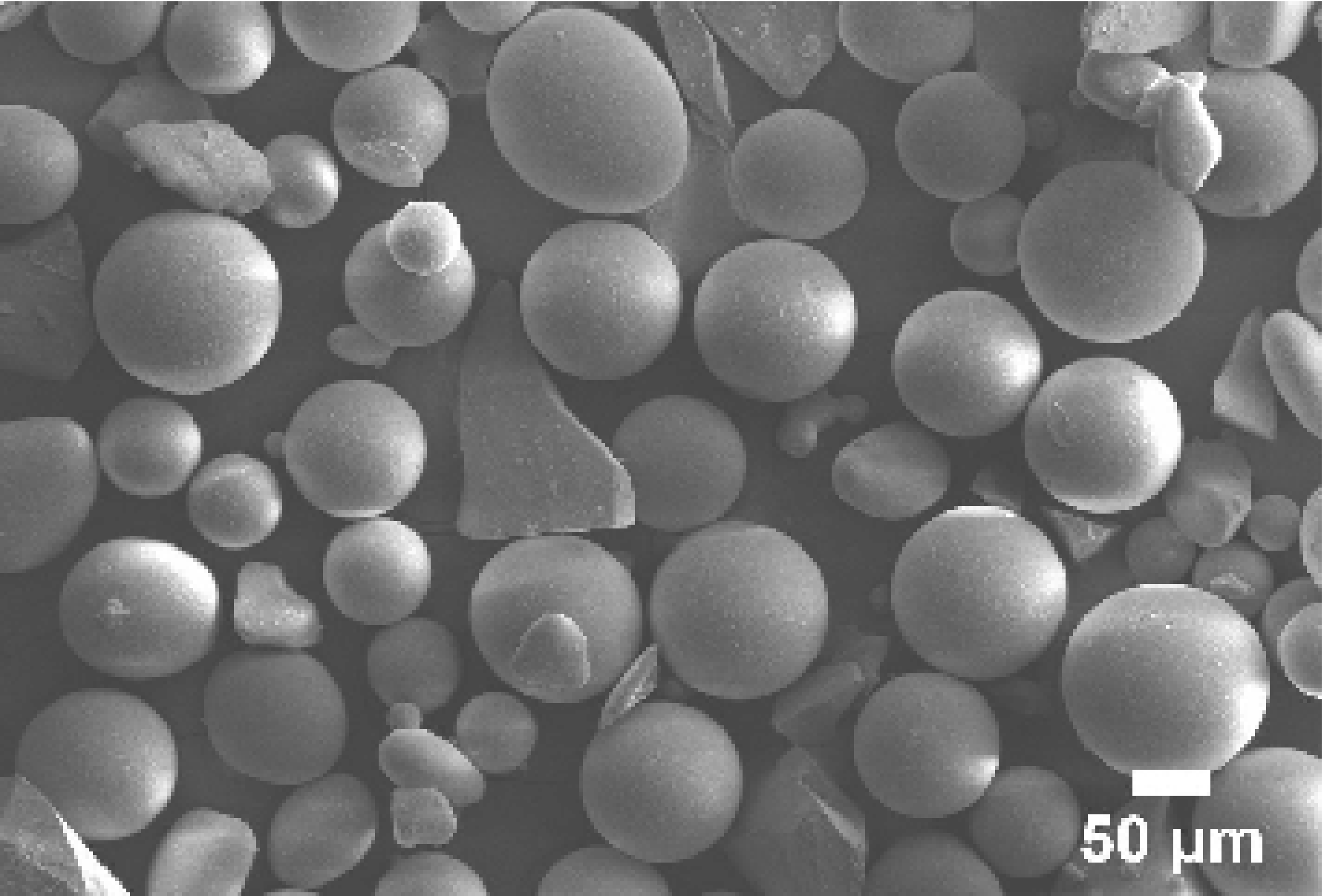}\\
      (d)
\end{tabular}
\end{minipage}
\caption{Grain images taken by SEM. (a) grain samples of 212 $-$ 300 $\mu m$ at 100 X. (b) individual grain at 2000 X, (c) grain samples of 106 $-$ 212 $\mu m$ at 300 X. (d) grain samples of 212 $-$ 300 $\mu m$ at 300 X.}
\label{fig:SEM}
\end{figure}

With the reservoir filled with grains, the exit valve was partially opened and the grains flowed through the tube. In order to evaluate the influence of the contact forces in the wave regime, we performed experiments with both smooth surface grains (beads used in only a few experiments) and rough surface grains (which passed many times through the glass tube and their surface was roughened by collisions with the walls and with other grains). The mass flow rate was determined from the time variation of the measured mass, using a chronometer and a balance with $\pm$ 0.01 g accuracy. Moreover, Scanning Electron Microscopy (SEM) was used to determine the median diameter of the grains $d_{50}$. The measured diameters were $d_{50}$ = 110 $\pm$ 3 $\mu m$ for the batch of grains with diameters between 106 and 212 $\mu m$, and $d_{50}$ = 225 $\pm$ 10 $\mu m$ for the batch of grains with diameters between 212 and 300 $\mu m$. As an example, Fig. \ref{fig:SEM} illustrates some images acquired by SEM, which shows some irregularly-shaped grains.

\begin{figure}[h!]
\begin{center}
\includegraphics[scale = 0.22, clip]{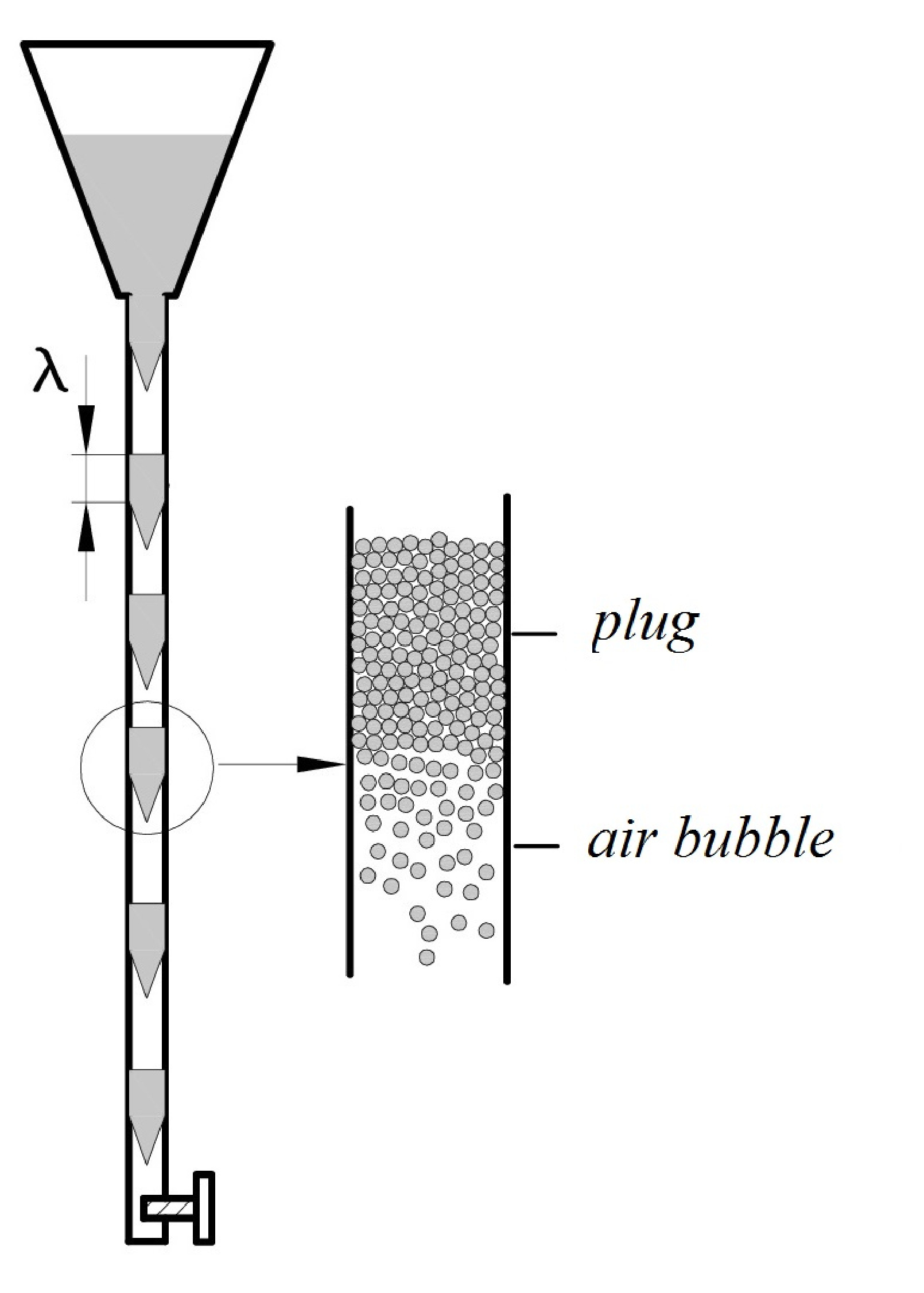}\\
\caption{Schematic view of the density waves, $\lambda$ is the length of the granular plug.}
\label{fig:waves}
\end{center}
\end{figure}

The grains falling through the transparent pipe were filmed with a high-speed camera acquiring at a maximum frequency of $1000\,Hz$ with a 1280 $px$ $\times$ 1024 $px$ spatial resolution. In order to provide the necessary light for low exposure times while avoiding beat frequencies between the light source and the camera frequency, LED (Low Emission Diode) lamps were branched to a continuous current source. For these experiments, the camera frequency was set to $250\,Hz$ and $500\,Hz$. The density waves were experimentally obtained with a moderate constriction at the bottom of the tube. Figure \ref{fig:waves} shows a schematic view of these waves, where the high- and low-compactness regions are designated as plugs and air bubbles, respectively, and $\lambda$ is the mean length of plugs.

\begin{figure}[h!]
\begin{center}
\includegraphics[scale = 0.30, clip]{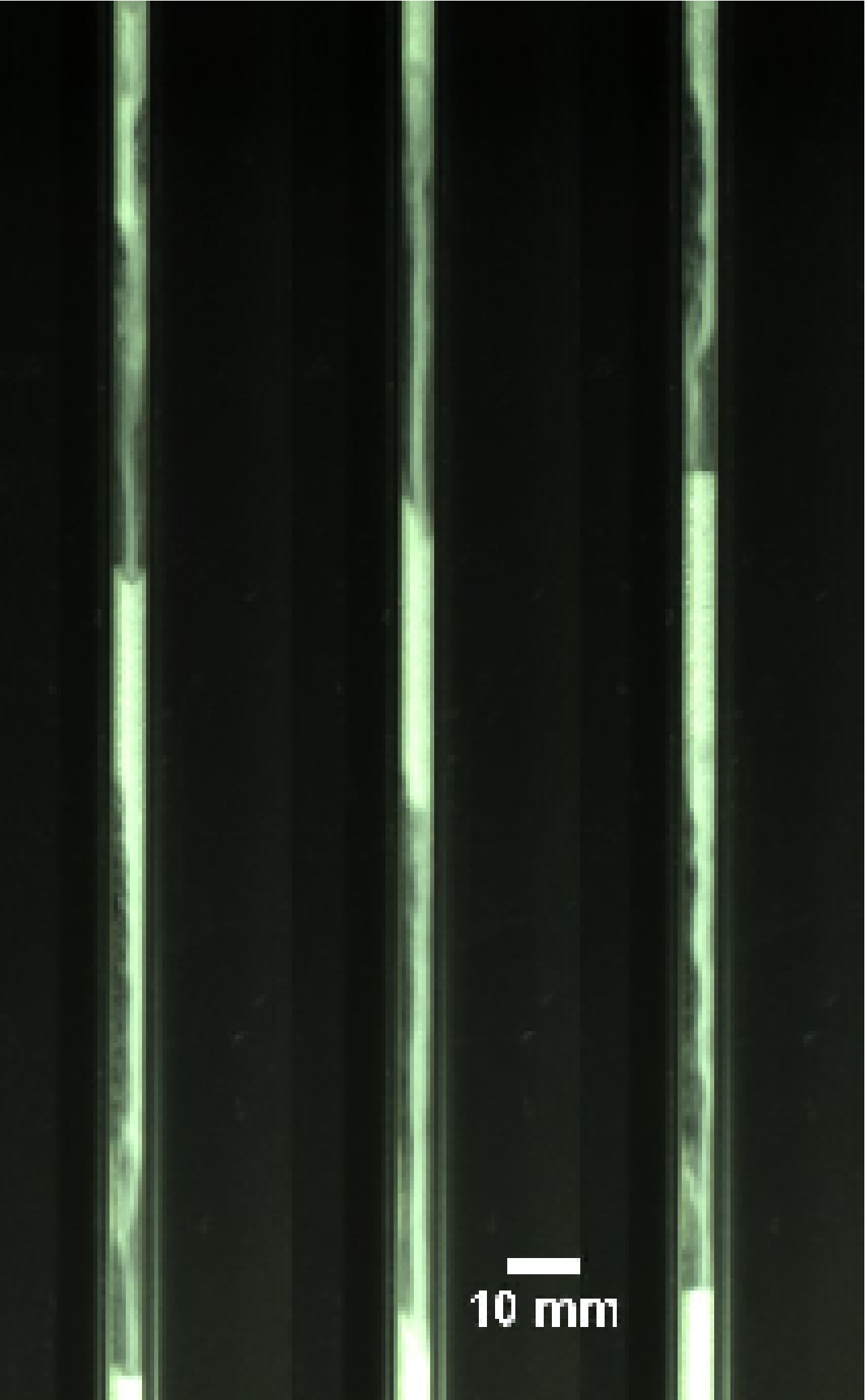}\\
\caption{Images acquired during the tests showing density waves with negative celerity. The time between frames is 0.06 $s$.}
\label{fig:image}
\end{center}
\end{figure}

A large number of images were acquired during the tests (in the order of thousands); therefore, we developed an image processing code in order to treat sequentially all the images obtained. The image processing code identifies the patterns in the RGB (red, green and blue) images, tracks them along images, and calculates the lengths and the celerities of the granular plugs. Figure \ref{fig:image} illustrates three frames of a test run for which the camera frequency was set to 250 $Hz$, but the time between these frames is 0.06 $s$. It was observed that the celerity of density waves was negative (contrary to granular flow direction). Subsection \ref{subsection_processing} describes briefly the image processing code developed in MatLab environment. 

\subsection{Digital image processing}
\label{subsection_processing}

An image processing code was developed during this work to automatically treat the images obtained. Therefore, MatLab scripts were written in order to automatically identify the granular plugs in the RGB images, to follow them along the images, and to compute the lengths and celerities of granular plugs. The following is a brief explanation of the code.

\begin{figure}[h!]
\begin{center}
\includegraphics[scale = 0.5, clip]{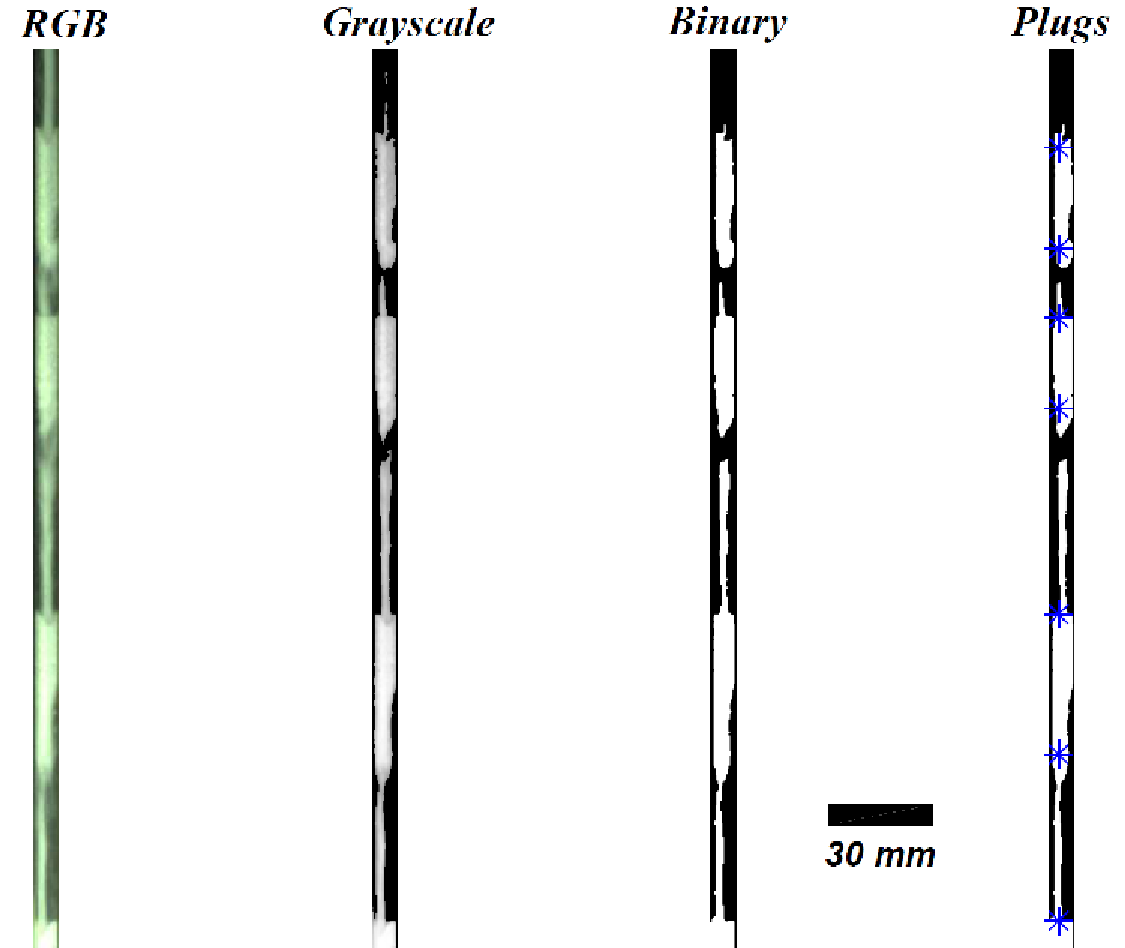}\\
\caption{Example of plugs identification by the code. From left to right: image in RGB format, image converted to grayscale, image converted to binary format, and identified the upper and bottom plug positions, marked with asterisks.}
\label{fig:code}
\end{center}
\end{figure}

Each image, which corresponds to a movie frame, is acquired in RGB format during the tests and it is saved as a matrix. The first steps of the code are to open sequentially each one of these matrices, and to load scaling and threshold information, related to the image and light calibrations (in order to convert pixels to mm, a calibration image was acquired before each test). Next, in each image the code detects the regions that contain the granular flow, and that is limited by the tube walls; therefore, the code only considers information belonging to this region. In the following, each image is converted to grayscale, and then to binary scale. Next, with the smallest granular plug assumed to scale with the tube diameter, the code searches the grain regions with lengths greater than $3\, mm$, and these are considered as the granular plugs. Once the plugs have been identified, the code saves their upper and bottom positions in a matrix. Finally, the mean celerity of plugs is determined by calculating the slope of their upper positions, the mean length of plugs is determined by calculating the difference between their upper and bottom positions, and spatiotemporal  diagrams were constructed. Figure \ref{fig:code} shows an example of plugs identified by the code for one image. From left to right, Fig. \ref{fig:code} shows the image in RGB format, the image converted to grayscale, the image converted to binary format, and the identified upper and bottom positions of the plugs, marked with asterisks. The main steps of the code are shown below in a schematic form. 

\begin{enumerate}
	\item Entry:
	\begin{enumerate}
		\item Image
		\item Frequency
		\item Scale
		\item Threshold
	\end{enumerate}
	\item Loop (until the total number of images is achieved):
	\begin{enumerate}
		\item Read image
		\item Select image area that contains the grain flows 
		\item Convert RGB image to grayscale image 
		\item Convert grayscale image to binary image
		\item Search areas greater than the tube diameter
		\item Show images while they are computing
		\item Store granular plug positions at different times
	\end{enumerate}
	\item Post-processing:
	\begin{enumerate}
		\item Computation of the mean celerity of granular plugs $v_p$ 
		\item Computation of the mean length of granular plugs $\lambda$
	\end{enumerate}
	\item Graphical display:
	\begin{enumerate}
		\item Plot the upper and bottom plug positions as a function of time (spatiotemporal diagram)
		\item Plot the mean length of plugs $\lambda$ as a function of mass flow rate $\dot{m}$
		\item Plot the mean celerity of plugs $v_{p}$ as a function of mass flow rate $\dot{m}$
	\end{enumerate}
\end{enumerate}

\section{Experimental results}

Density waves were observed as soon as the granular flow began. The density waves had positive or negative mean celerity, and they sometimes presented an oscillatory component. Aider et al. \cite{Aider} proposed that different celerity behaviors are due to different granular flow rates and humidity. 

Two wave flow regimes were identified based on spatiotemporal  diagrams. The first corresponding to density waves that propagated at a constant celerity (known as propagative wave regime), and the second corresponding to oscillating waves, with low and large amplitude oscillations. Previous works \cite{Aider, Bertho_1} have reported that density waves appeared when the mass flow rate $\dot{m}$ was within 1.5 $-$ 5 $g/s$, (with plug compactness typically within 55$-$60$\%$), separated by internal air bubbles, in which the grain dynamics can be modeled locally as a free fall motion. In the present work, we observed density waves for the $\dot{m}$ values ten times smaller than those previously reported, and slightly smaller relative humidity values than those previously reported.

In our opinion, the relatively small mass flow rates observed in our study are due to two factors. The first is associated to the relative humidity in our tests. We observed density waves at relative humidity values smaller than those reported in previous works. In addition, in our experiments the relative humidity was continuously controlled, and this parameter was kept constant during each test (Tabs. \ref{tab:waves1}, \ref{tab:waves2}). The second is associated to the geometry and superficial roughness of the grains used. The images taken by SEM allowed to observe that some grains were not perfect spheres (Figs. \ref{fig:SEM}c, \ref{fig:SEM}d), and that some grains contained little incrustations on their surface (Fig. \ref{fig:SEM}b).

In order to determine the mean celerity $v_p$ and the mean length $\lambda$ of granular plugs, we constructed spatiotemporal diagrams. These diagrams show the plug positions as a function of time; therefore, $v_p$ can be obtained from the mean slope of the plug positions, and $\lambda$ can be obtained by subtracting the upper and bottom positions of each plug.

\subsection{Propagative wave regime}

\begin{figure}
\begin{minipage}[c]{\textwidth}
\centering
\begin{tabular}{c}
\includegraphics[scale=0.65]{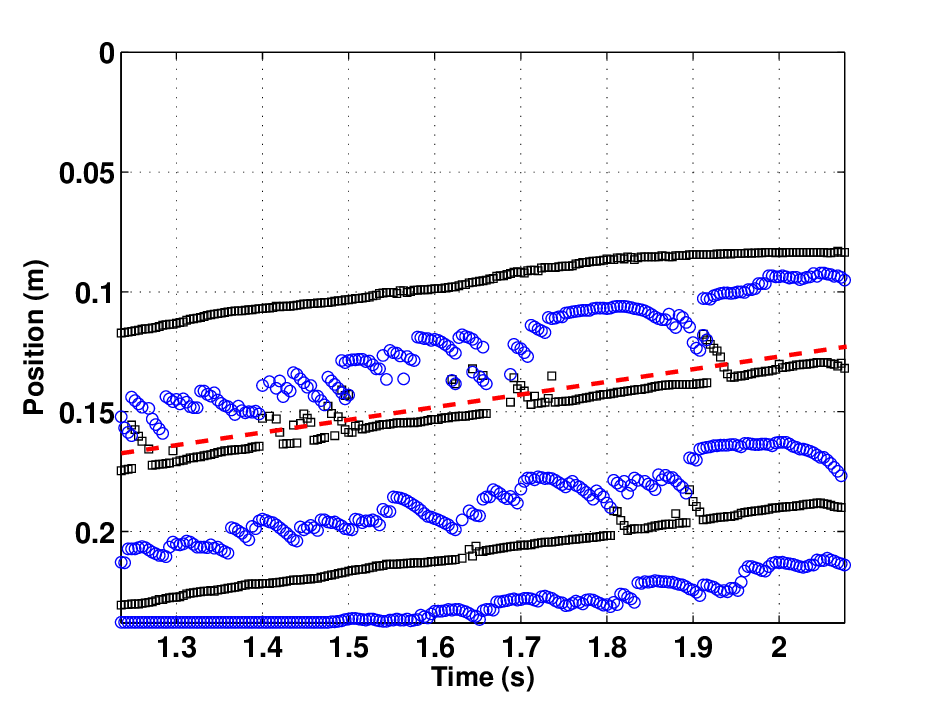}\\
      (a)
\end{tabular}
\end{minipage} \hfill
\begin{minipage}[c]{\textwidth}
\centering
\begin{tabular}{c}
\includegraphics[scale=0.65]{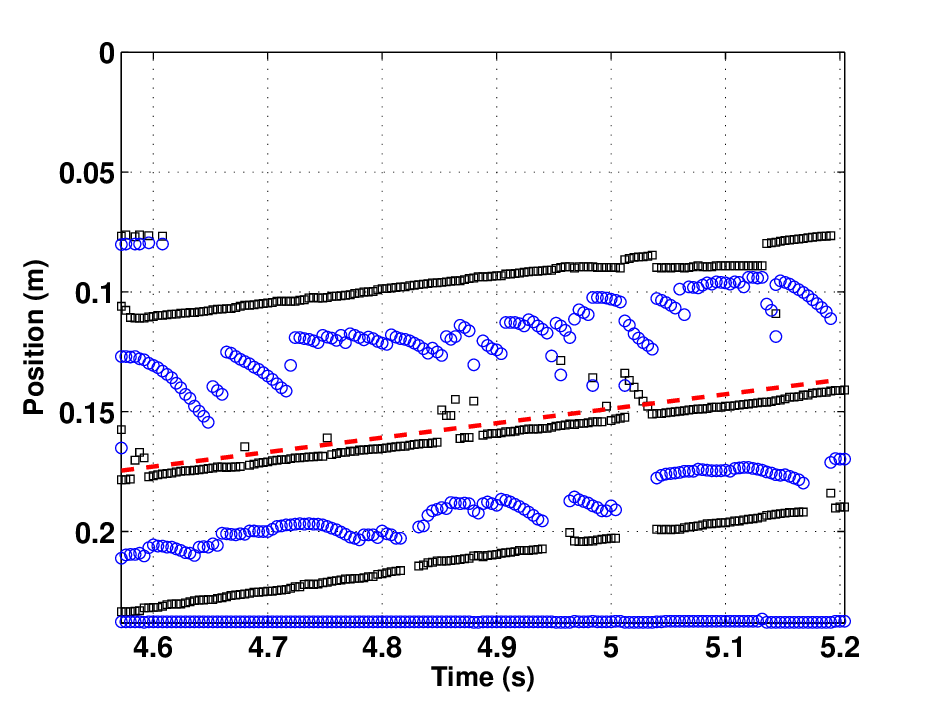}\\
      (b)
\end{tabular}
\end{minipage}
\caption{Spatiotemporal diagrams corresponding to propagative wave regime for rough surface grains. The upper and bottom plug positions are shown by black squares and blue circles, respectively. The mean celerity of plugs $v_p$ is shown in the dashed line, $v_{p}$ = -0.0538 m/s, $\dot{m}$ = 0.55 g/s.}
\label{fig:propag_2}	
\end{figure}

This flow regime appeared with smooth and rough surface grains, and diameters within 106 and 212 $\mu m$. In all tests within this diameter range, the density waves propagated upward, i.e., contrary to flow direction, which according to our coordinate reference system represented a negative celerity. Figure \ref{fig:propag_2} displays two spatiotemporal diagrams in the case of rough surface grains. The diagrams in Fig. \ref{fig:propag_2} are representative of all tests with grains in this diameter range.

The vertical axis corresponds to the distances measured along the pipe, increasing from top to bottom, and the horizontal axis corresponds to time, increasing from left to right. The height of the field of view is 237 $mm$. The top of this figure is located at 30 $cm$ downstream of the tube's upper end. The literature reports that between 7 and 20 $cm$ is required for fully developed plugs to build up \cite{Aider, Poschel}. The variations as functions of time of the upper and bottom plug positions are shown by black squares and blue circles, respectively. The mean celerity of plugs $v_p$, can be determined by the slope obtained for the upper zone. The slope is constant for a plug and also from one plug to another, then $v_p$ can be easily measured. In addition, $v_p$ varies as a function of the mass flow rate $\dot{m}$, $\dot{m}$ = 0.55 $g/s$ for Fig. \ref{fig:propag_2}.

\begin{figure}[h!]
\begin{center}
\includegraphics[scale = 0.65,clip]{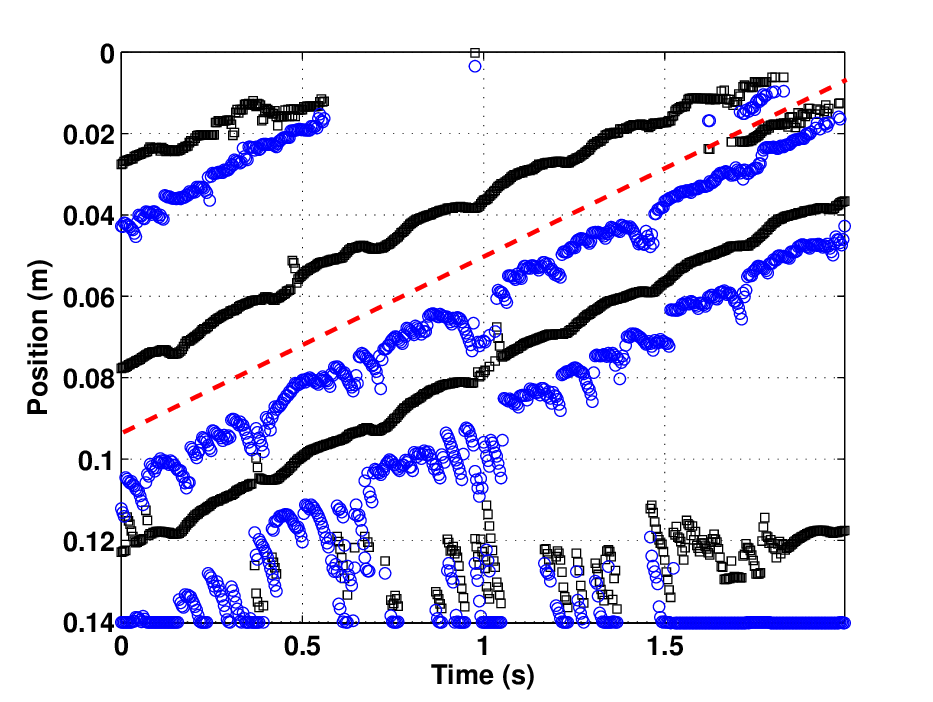}\\
\end {center}
\caption{Spatiotemporal diagram corresponding to propagative wave regime for smooth surface grains. The upper and bottom plug positions are shown by black squares and blue circles, respectively. The mean celerity of plugs $v_p$ is shown in the dashed line, $v_{p}$ = -0.0408 m/s, $\dot{m}$ = 0.40 g/s.}
\label{fig:propag_nov}	
\end {figure}

Figure \ref{fig:propag_nov} presents a spatiotemporal diagram for which the top is located at 45 $cm$ downstream of the tube's upper end, for the case of smooth surface grains. From Figs \ref{fig:propag_2} and \ref{fig:propag_nov}, we conclude that $v_p$ is somewhat independent of the vertical position of the analyzed field, i.e., all captured plugs inside the diagrams have the same mean celerity $v_p$; therefore, we can assert that $v_p$ is constant along the tube. This is in agreement with Aider et al. \cite{Aider}, who also reported that $v_p$ is constant once a stationary propagation regime has been reached. In order to verify if the observed plugs were moving in a stationary regime, we analyzed spatiotemporal diagrams constructed over different time intervals during each test run. The two diagrams presented in Fig. \ref{fig:propag_2} belong to the same test. The test starts in Fig. \ref{fig:propag_2}a, the duration is almost one second, and after 2.4 seconds (not shown in Fig. \ref{fig:propag_2}), the test continues in Fig \ref{fig:propag_2}b, where the plugs maintain the same celerity they had before. 

Spatiotemporal diagrams using different time intervals for the same test runs were constructed for both rough and smooth surface grains. The results were always similar, with granular plugs propagating upward with a constant mean celerity $v_p$, and the mean celerity was negative for all the tests, i.e., the plugs always propagated upward, contrary to the flow direction. In previous works \cite{Aider, Bertho_1, Raafat}, the authors reported that the plugs propagated downward, in same direction of the flow. This difference may be related to the different environmental conditions and the mass flow rates used in our tests. In the studies mentioned in this paper, the mass flow rates were ten times greater than our rates, and the relative humidity was slightly larger.

\begin{figure}[h!]
\begin{center}
\includegraphics[scale = 0.12,clip]{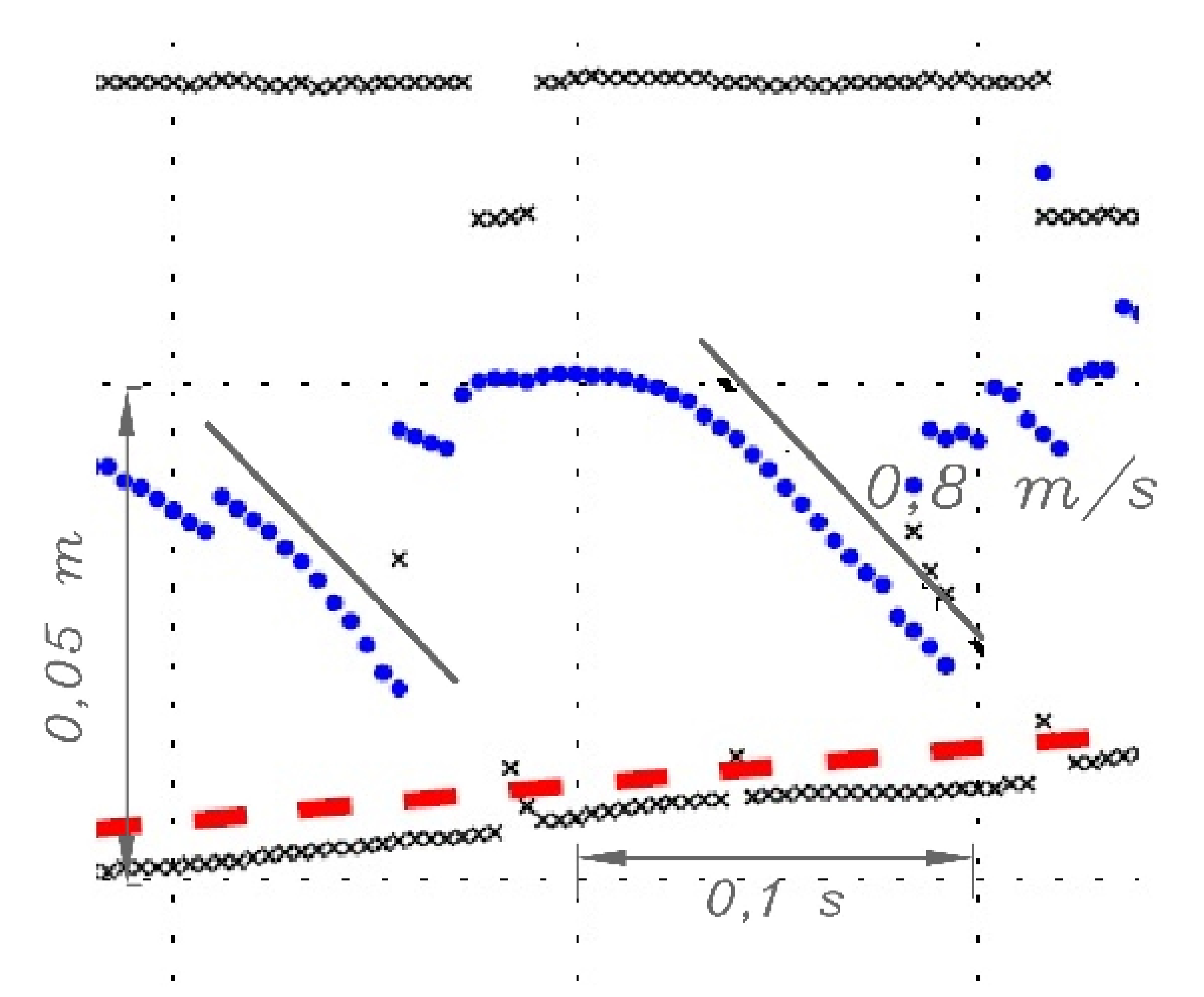}\\
\end {center}
\caption{Enlarged view of a spatiotemporal diagram. The mean celerity of plugs $v_{p}$ = -0.0366 m/s and the mass flow rate $\dot{m}$ = 0.37 g/s.}
\label{fig:vista}	
\end {figure}

The spatiotemporal diagrams in Figs. \ref{fig:propag_2} and \ref{fig:propag_nov} show that the bottom of plugs, identified by blue circles in the diagrams, exhibit strong curvatures followed by stronger slopes in some regions. This suggests that in these regions some grains are accelerated by gravity as they fall out of a plug, and into the bubble immediately below with a free fall motion. The curvature then gives the accelerating transient and the stronger slope the free fall velocity. The grains then leave the bubble, and are slowed down when they reach the upper boundary of the next plug. Figure \ref{fig:vista} shows an enlarged view of a spatiotemporal diagram for which $v_{p}$ = -0.0366 m/s and $\dot{m}$ = 0.37 g/s. In this figure, the strong curvatures followed by stronger constant slopes (continuous lines drawn in Fig. \ref{fig:vista}) are highlighted. From the constant slope lines, the grains' velocity was computed to be in the order of 0.8 $m/s$, which corresponds to the free fall velocity in the bubble region. For comparison, we estimated the final velocity by assuming free fall in the bubble and neglecting the air resistance. This gives approximately 0.9 $m/s$, which is in agreement with the velocity obtained from the spatiotemporal diagram.

\begin{figure}[h!]
\begin{center}
\includegraphics[scale = 0.65,clip]{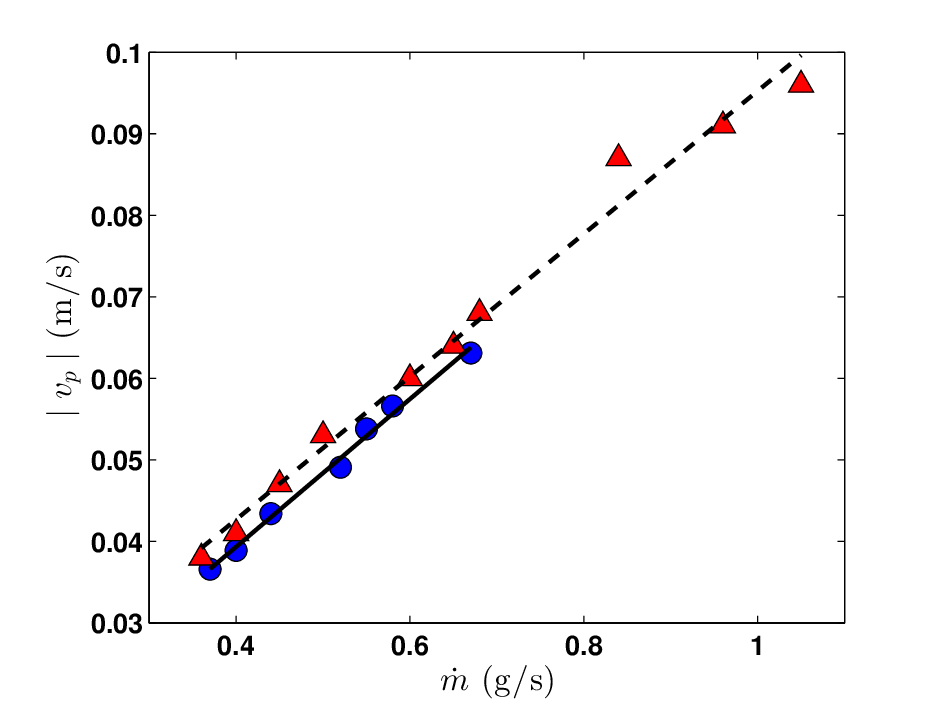}\\
\end{center}
\caption {Variation of mean celerity of plugs $v_{p}$ as a function of mass flow rate $\dot{m}$ in the propagative wave regime. The symbols $\bigcirc$ and $\triangle$ correspond to tests using rough grains and smooth grains, respectively, and the continuous and dashed lines represent linear fittings.}
\label{fig:vpm} 
\end{figure}

\begin{figure}[h!]
\begin{center}
\includegraphics[scale = 0.65,clip]{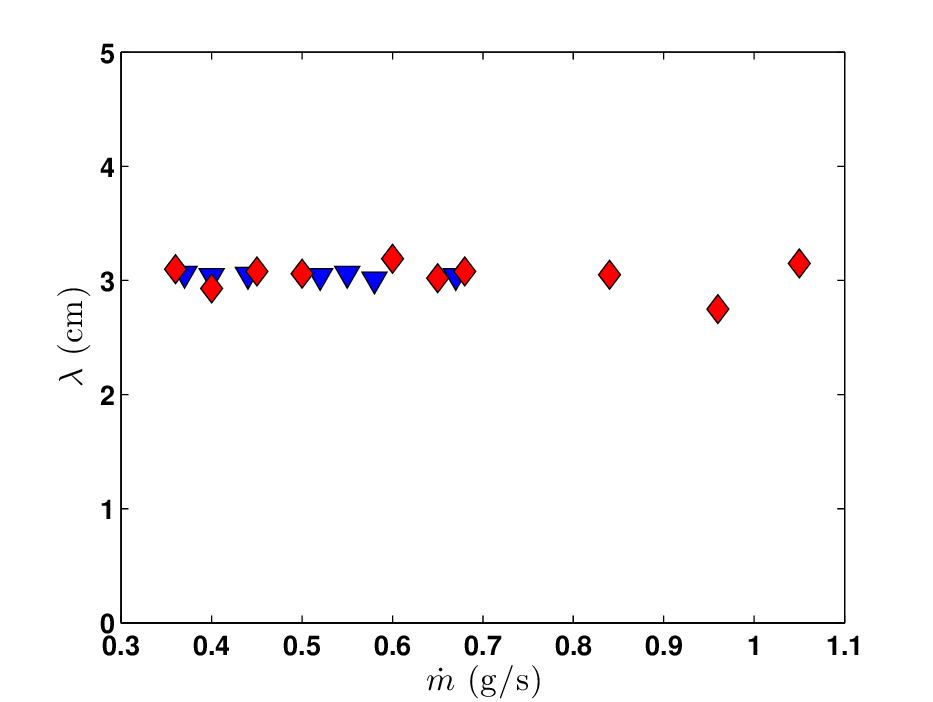}\\
\end{center}
\caption {Mean length of plugs $\lambda$ as a function of mass flow rate $\dot{m}$ in the propagative wave regime. The symbols $\Diamond$ and $\bigtriangledown$ correspond to the tests using smooth grains and rough grains, respectively. The size of the plugs is approximately 30 $mm$ ($\lambda$ $\approx$ 10$D$).}
\label{fig:lbm} 	
\end{figure}

Table \ref{tab:waves1} illustrates the main results obtained for the propagative wave regime. For each test run, Tab. \ref{tab:waves1} presents the mass flow rate $\dot{m}$, the room relative humidity $H$, the surface state of grains, the mean celerity $v_p$, and the mean length of plugs $\lambda$. The results of Tab. \ref{tab:waves1} are presented in Figs. \ref{fig:vpm} and \ref{fig:lbm} in order to investigate the relation between the mean celerity of plugs and the mass flow rate, and between the mean length of plugs and the mass flow rate. Previous works have shown that the  mean celerity of plugs has a linear dependence on the mass flow rate \cite{Aider, Bertho_1, Raafat}. Therefore, we investigated if there would be linear dependence in our experiments, despite the upward motion of plugs.

\begin{table}[ht!]
	\begin{center}
	\begin{tabular}{c c c l c c}
	\hline\hline
			run & $\dot{m}$ & $H$ & surface & $v_{p}$ & $\lambda$ \\
			$\cdots$ & $g/s$ & $\%$ & $\cdots$ & $ m/s$ & $cm$  \\
\hline
			1 & 0.37 & 44.1 & rough & -0.0366 & 3.06\\
			2 & 0.52 & 41.4 & rough & -0.0491 & 3.04\\
			3 & 0.55 & 41.3 & rough & -0.0538 & 3.06\\
			4 & 0.40 & 41.6 & rough & -0.0389 & 3.04\\
			5 & 0.67 & 42.3 & rough & -0.0631 & 3.04\\
			6 & 0.44 & 43.0 & rough & -0.0434 & 3.05\\
			7 & 0.58 & 42.1 & rough & -0.0566 & 3.01\\
			8 & 0.66 & 64.8 & smooth  & -0.0681 & 3.20\\
			9 & 0.36 & 59.2 & smooth  & -0.0382 & 3.10\\
			10 & 0.84 & 53.5 & smooth  & -0.0873 & 3.05\\
			11 & 0.40 & 62.9 & smooth  & -0.0408 & 2.93\\
			12 & 1.05 & 63.6 & smooth  & -0.0963 & 3.15\\
		    13 & 0.50 & 58.2 & smooth  & -0.0530 & 3.06\\
			14 & 0.68 & 56.5 & smooth  & -0.0682 & 3.08\\
			15 & 0.45 & 60.9 & smooth  & -0.0470 & 3.08\\
			16 & 0.65 & 57.5 & smooth  & -0.0641 & 3.02\\
			17 & 0.96 & 52.5 & smooth & -0.0914 & 2.75\\
\hline
		\end{tabular}
			\caption{$ \dot{m} $, $ H $, surface state, $v_{p}$, $\lambda$ for each test run in the propagative wave regime}
		\label{tab:waves1}
	\end{center}
\end{table}

Figure \ref{fig:vpm} presents the variation of $v_p$ (absolute values) as a function of $\dot{m}$. The celerity of plugs increases linearly with $\dot{m}$ and ranges between -0.0366 and -0.0963 $m/s$. The linear dependency of $v_p$ with $\dot{m}$ is in agreement with that reported in the literature \cite{Aider, Bertho_1, Raafat}. In addition, Fig. \ref{fig:vpm} shows that the celerity of smooth-grained plugs is larger than the celerity of coarse-grained plugs.

Figure \ref{fig:lbm} presents the length of plugs $\lambda$ as a function of the mass flow rate $\dot{m}$, and it shows that $\lambda$ is almost independent of $\dot{m}$. The mean size of plugs obtained from the spatiotemporal diagrams was $\lambda$ $\approx$ 10$D$. In addition, the length of the plugs appears to be independent of the surface state of the grains.

\subsection{Oscillating wave regime}

The oscillating regime appeared when the experiments were conducted using a batch of rough-surface glass spheres and diameters ranging between 212 and 300 $\mu m$. The mass flow rates were between 0.25 and 0.4 $g/s$. In this regime, it was observed that the density waves oscillated over a non-zero drift celerity, which can be considered as the mean celerity of plugs $v_p$. The plugs always propagated downward, in the same direction as the granular flow, and $v_p$ was constant along the tube. 

\begin{figure}
\begin{minipage}[c]{\textwidth}
\centering
\begin{tabular}{c}
\includegraphics[scale=0.65]{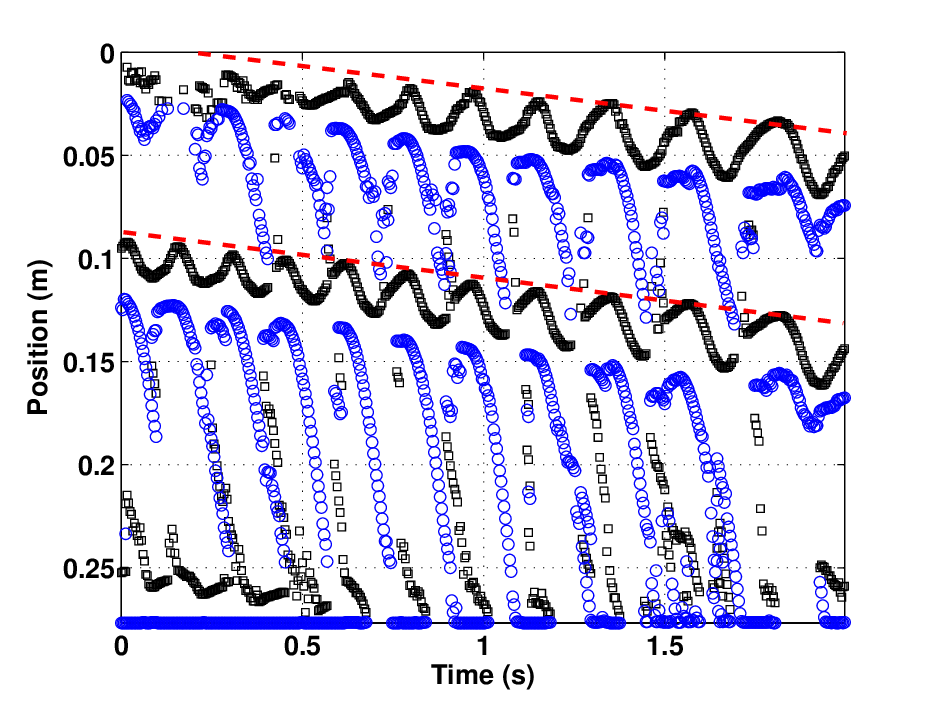}\\
      (a)
\end{tabular}
\end{minipage} \hfill
\begin{minipage}[c]{\textwidth}
\centering
\begin{tabular}{c}
\includegraphics[scale=0.65]{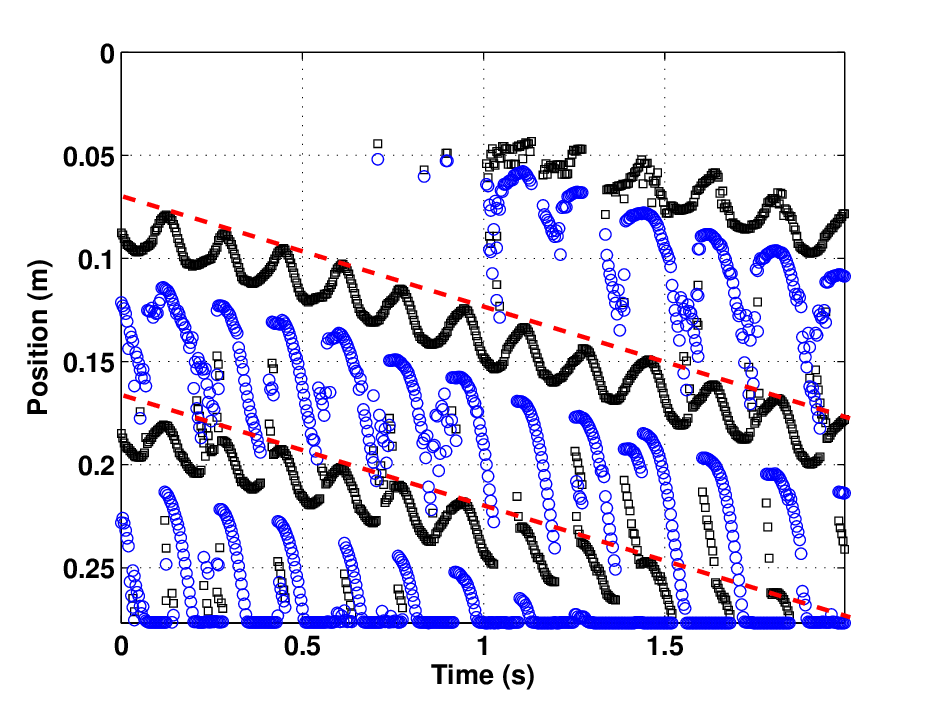}\\
      (b)
\end{tabular}
\end{minipage}
\caption {Spatiotemporal diagram of low amplitude oscillating waves. The upper and bottom plug positions are shown by black squares and blue circles, respectively. The dashed line corresponds to the mean drift celerity $v_p$. (a) $v_p$ = 0.0250 $m/s$, $\dot{m}$ =  0.26 $g/s$. (b) $v_p$ = 0.0501 $m/s$, $\dot{m}$ =  0.37 $g/s$. The frequency of oscillations is in the order of 7 $Hz$.}
\label{fig:small} 
\end{figure}

Density waves oscillating at low amplitude appeared in some tests. Figure \ref{fig:small} displays spatiotemporal diagrams with these characteristics. The frequency of the observed oscillations was 7 $Hz$. Figures \ref{fig:small}a and \ref{fig:small}b correspond to the diagrams obtained with values very close to the superior and inferior limits of $\dot{m}$ determined for this wave regime. These figures show that the oscillating frequency is independent of the mass flow rate.   

Table \ref{tab:waves2} illustrates the values of the mass flow rate $\dot{m}$, the room relative humidity $H$, the mean celerity $v_p$, and the mean length of plugs $\lambda$, for each test run, measured for density waves oscillating at low amplitude.

\begin{table}[ht!]
\begin{center}
\begin{tabular}{c c c c c}
\hline\hline
			run & $\dot{m}$ & $H$ & $v_{p}$ & $\lambda$ \\
			$\cdots$ & $g/s$ & $\%$ & $ m/s$ & $cm$  \\
\hline
			1 & 0.35 & 60.7 & 0.0401 & 3.41\\
			2 & 0.37 & 52.7 & 0.0501 & 3.15\\
			3 & 0.41 & 52.8 & 0.0652 & 3.36\\
			4 & 0.26 & 52.0 & 0.0250 & 3.33\\
			5 & 0.26 & 59.2 & 0.0200 & 3.05\\
			6 & 0.25 & 59.0 & 0.0232 & 3.25\\
			7 & 0.26 & 58.8 & 0.0220 & 3.21\\
			8 & 0.30 & 61.5 & 0.0291 & 2.98\\
			9 & 0.35 & 62.1 & 0.0420 & 3.30\\
		   10 & 0.33 & 60.7 & 0.0370 & 3.25\\
		   11 & 0.39 & 54.9 & 0.0550 & 3.15\\
		   12 & 0.31 & 62.3 & 0.0334 & 3.15\\
		   13 & 0.36 & 57.3 & 0.0420 & 3.16\\
		   14 & 0.25 & 61.5 & 0.0289 & 3.15\\ 
		   15 & 0.27 & 52.5 & 0.0263 & 3.23\\
\hline
\end{tabular}
\caption{$ \dot{m} $, $ H $, $v_{p}$, $\lambda$ for each test run, in the case of density waves oscillating at low amplitude}
\label{tab:waves2}
\end{center}
\end{table}

\begin{figure}[ht!]
\begin{center}
\includegraphics[scale = 0.65,clip]{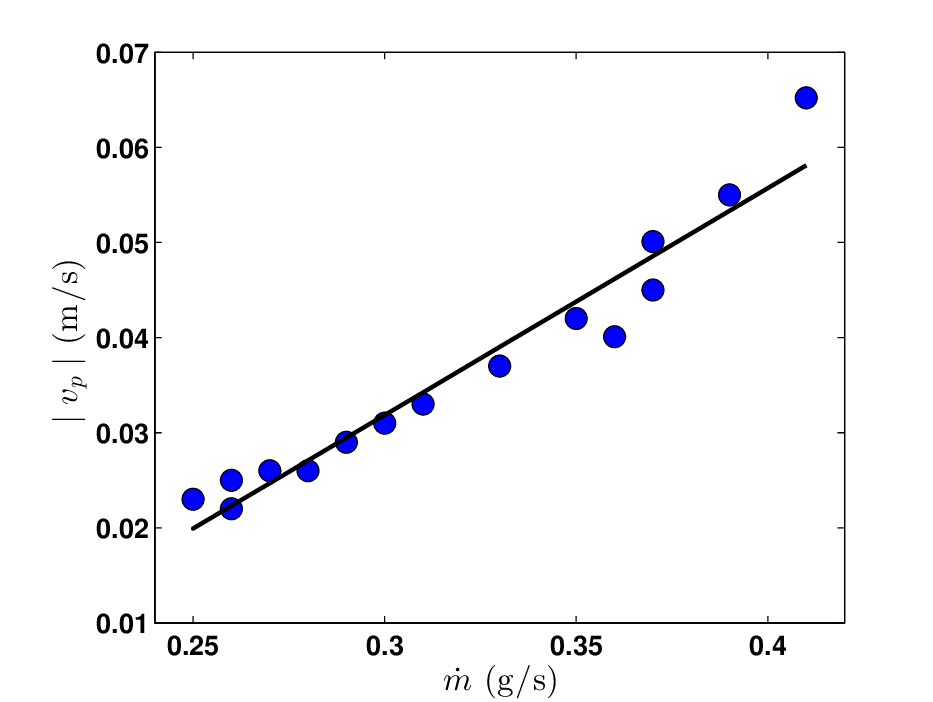}\\
\end{center}
\caption {Mean celerity of plugs $v_{p}$ as a function of mass flow rate $\dot{m}$ in the oscillating wave regime. The continuous line corresponds to a linear fitting.}
\label{fig:vpm_p} 
\end{figure}

\begin{figure}[ht!]
\begin{center}
\includegraphics[scale = 0.65,clip]{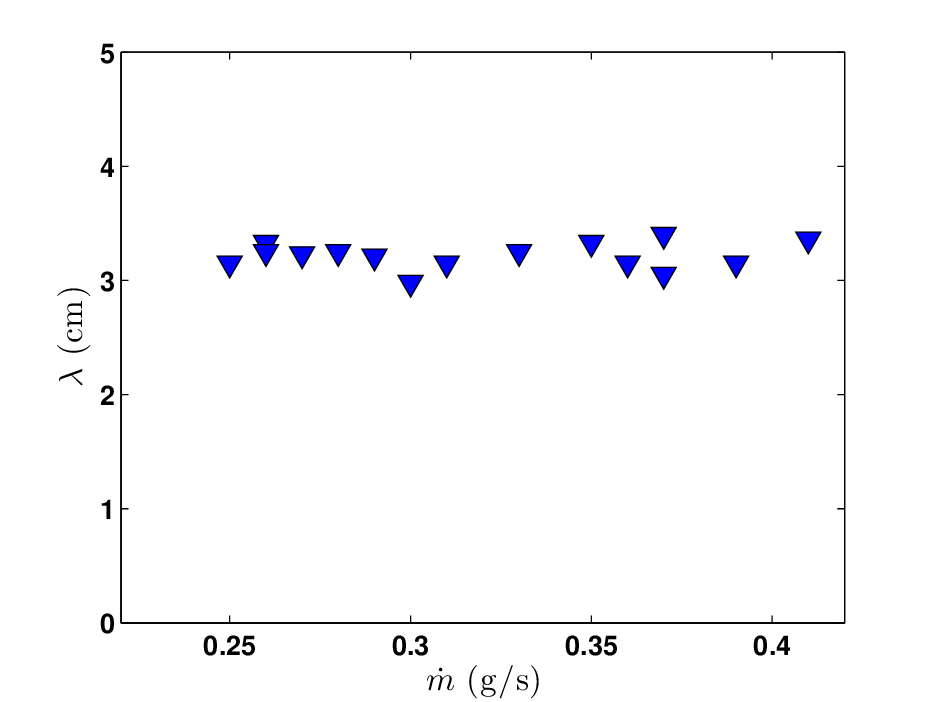}\\
\end{center}
\caption {Mean length of plugs $\lambda$ as a function of mass flow rate $\dot{m}$ in the oscillating wave regime. The size of the plugs is of approximately 30 $mm$ ($\lambda$ $\approx$ 10$D$).}
\label{fig:lbm_p} 
\end{figure}

The results of Tab. \ref{tab:waves2} are shown in Figs. \ref{fig:vpm_p} and \ref{fig:lbm_p} in order to investigate the relation between the mean celerity of plugs and the mass flow rate, and between the mean length of plugs and the mass flow rate. Figure \ref{fig:vpm_p} presents the mean celerity of plugs $v_{p}$ as a function of mass flow rate $\dot{m}$ for the low amplitude oscillating waves, and shows that there is an approximate linear dependence of $v_p$ on $\dot{m}$. The range of $v_p$ values is within 0.2 $-$ 0.65 $m/s$, which is considerably lower than previously reported results; however, the linear dependence remains. Figure \ref{fig:lbm_p} displays the mean length of plugs $\lambda$ as a function of mass flow rate $\dot{m}$ in the oscillating wave regime, and shows that $\lambda$ is independent of the mass flow rate, with its value of approximately 10$D$. We note that the propagative regime presented identical results; therefore, we conclude that the mean length of plugs is independent of both the mass flow rate and the wave regime.

The granular flow surrounded by air is a two phase flow system, and some analogies with other two phase systems can be explored, as, for example, air-water flows. These systems are characterized by high compressibility (due to air bubbles) and high density ratios (due to mean density of the plugs compared to air). The sound velocity computed for those systems is generally low, in the order of a few $m/s$, resulting in low resonance frequencies. Using some approximations, it is possible to calculate a resonance frequency equivalent to 4 $Hz$ in this case. This value is in accordance with the frequency calculated only using the spatiotemporal diagrams (7 $Hz$), indicating that the mechanism of acoustic resonance may be related to the plugs length.

\begin{figure}[h!]
\begin{center} 
\includegraphics[scale=0.65]{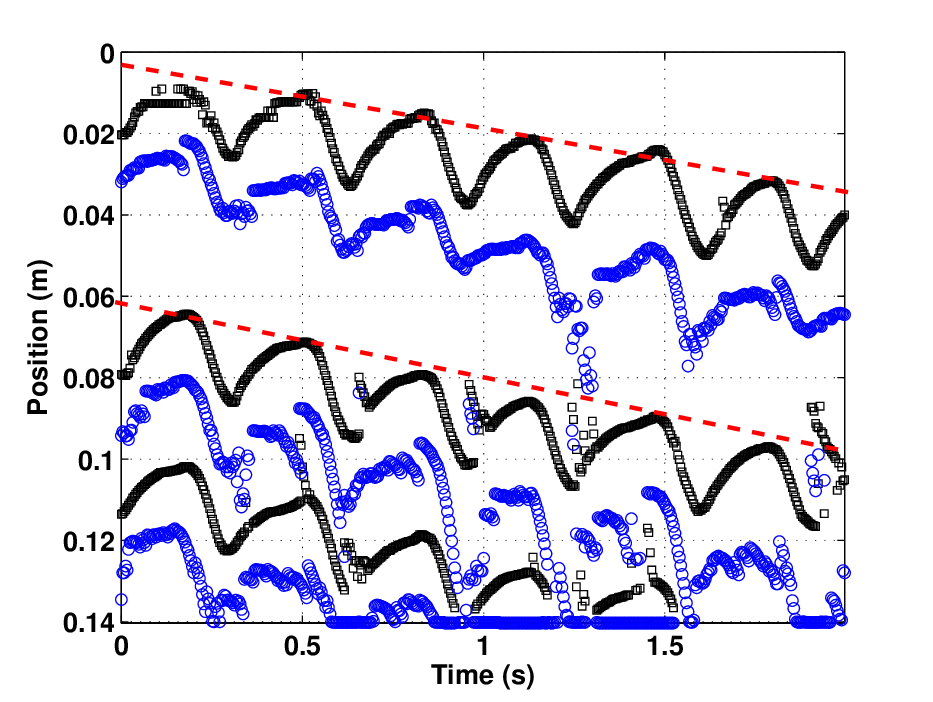}\\
\end{center}
\caption {Spatiotemporal diagram of large amplitude oscillating waves. Black squares and blue circles correspond to the upper and bottom plug positions, respectively. The dashed line corresponds to the mean drift celerity $v_p$, and the mean values are $v_p$ = 0.0190 $m/s$, $\dot{m}$ =  0.25 $g/s$. The frequency of oscillations is in the order of 4 $Hz$.}
\label{fig:large} 
\end{figure}

In the oscillating wave regime, oscillating flows with amplitudes slightly larger than those observed in Fig. \ref{fig:small} were also detected. The frequency of the waves was in the order of 4 $Hz$. Figure \ref{fig:large} shows a typical spatiotemporal diagram corresponding to this situation, which shows that the mean celerity of plugs $v_p$, represented by the dashed line, is constant and positive (plugs propagate downward). 

The amplitude of the oscillations suggests that the granular plugs maintain static contact with the tube walls for some time, and afterward they start to move again, in a typical stick-slip movement. The friction forces between the plugs and the tube walls must be large enough to balance the weight of the grains and cause the plugs to stop for some time. Under these conditions, there is a redirection of forces within the plug, and the Janssen effect \cite{Janssen} is expected. Franklin and Alvarez \cite{Franklin_7}, using the Janssen's theory, calculated the mean size of a compact plug and reported the length of plugs was in the range $3 < \lambda/D < 11$, which is in accordance with the results of the present study.

\clearpage

\section{Gravity-driven flow of grains through pipes: a one-dimensional model}
\label{section:model}

\begin{figure} [h!]
\begin{center}
\includegraphics[scale=0.65]{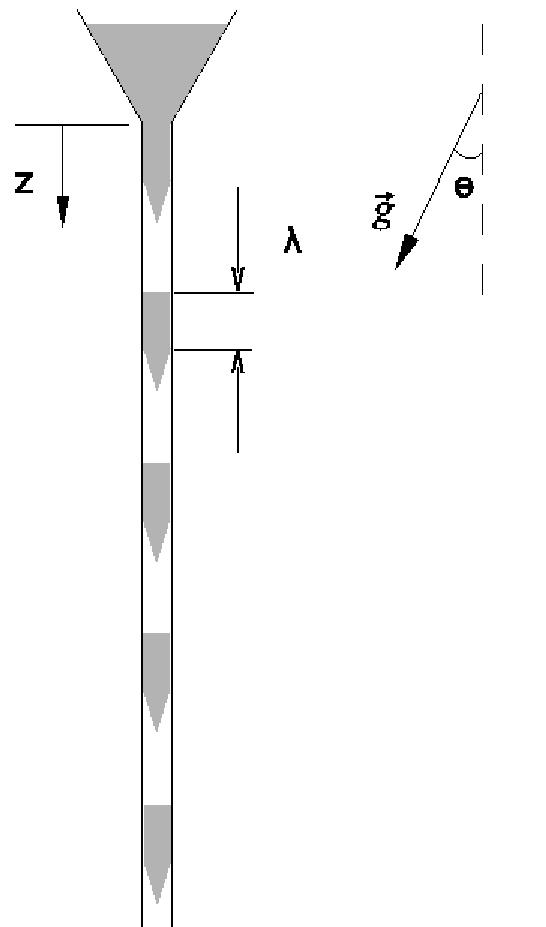}
 \caption{Layout of the gravitational granular flow through a narrow pipe. $z$ is the vertical coordinate, $\lambda$ is the length of the granular plugs, and $\theta$ is the pipe inclination with respect to the gravitational acceleration $\vec{g}$.}
\label{fig:escopo}
\end{center}
\end{figure}

The analyzed problem consists of cohesionless fine grains falling from a hopper through a narrow tube. The ratio between the tube diameter and the mean grain diameter is within $6 \leq D/d \leq 30$, the humidity is within $35<H<75\%$, and the grain size and specific mass are such that the air effects are not negligible. The tube is in a vertical (or almost vertical) position, i.e., $-10^o\,\leq\, \theta \,\leq\,10^o$, where $\theta$ is the pipe inclination with respect to the gravitational acceleration. Within this scope, density waves consisting of alternating high- and low- compactness regions are expected \cite{Aider,Bertho_1}. In the high-concentration regions, which are plugs of granular material, the compactness varies but is close to its maximum value, and grains in the plug periphery are in contact with the tube wall. Therefore, there is a redirection of forces within the plug and the Janssen effect is expected if the plugs are long enough \cite{Duran,Cambau}. In the low-concentration regions, which are air bubbles with dispersed free-falling grains, the air pressure increases owing to the stresses caused by the neighboring plugs as well as the volume decrease caused by the free-falling grains. Figure \ref{fig:escopo} shows the layout of the gravitational granular flow.

A one-dimensional model is proposed for this problem. The model consists of an equation of motion for the grains in a compact plug (or a compact regime), an air pressure equation, and a mass conservation equation of the grains. These equations, displayed below, are used in the stability analysis in Section \ref{section:analysis}. The stability analysis is performed to determine the length scale of the granular plugs.

\subsection{Mass conservation of the grains}

The mass transport equation of the grains is given by Eq. \ref{eq:mass1}. 

\begin{equation}
\frac{\partial c}{\partial t} + c \frac{\partial v_s}{\partial z} + v_s\frac{\partial c}{\partial z} = 0
\label{eq:mass1}
\end{equation}   

\noindent where $c$ is the compactness of granular plugs, $v_s$ is the local grain velocity, $z$ is the vertical coordinate, and $t$ is the time. Normalizing Eq. \ref{eq:mass1} by the characteristic length $L_c=D$, time $t_c=\sqrt{D/g}$, and velocity $v_c=\sqrt{gD}$, we obtain Eq. \ref{eq:mass2}:

\begin{equation}
\frac{\partial c}{\partial t^*} + c \frac{\partial v_{s}^*}{\partial z^*}+ v_{s}^*\frac{\partial c}{\partial z^*} = 0
\label{eq:mass2}
\end{equation} 

\noindent where $z^*=z/D$, $t^*=t/\sqrt{D/g}$, ${v_s}^*=v_s/\sqrt{gD}$. 

\subsection{Granular motion}
\label{section_granular_motion}

The equation of motion for the grains in compact regime is given by a balance between the grains acceleration, the friction between the grains and the tube wall, the forces due to air pressure and granular tension distribution, and the weight. This balance is given by Eq. \ref{eq:motion_ant},

\begin{equation}
\rho_{s}\left( \frac{\partial cv_{s}}{\partial t} + \frac{\partial c v_{s}^{2}}{\partial z} \right) = \rho_{s} \, cg\, \cos\, \theta - \frac{\partial P}{\partial z}  - \frac{\partial \sigma_{zz}}{\partial z} - \frac{4}{D}\sigma_{zr}
\label{eq:motion_ant}
\end{equation}

\noindent where $\rho_s$ is the specific mass of each grain, $g$ is the gravitational acceleration, $P$ is the air pressure, $\sigma_{zz}$ is the vertical stress operating on the grains, and $\sigma_{zr}$ is the stress between the tube wall and the grains. We use here the closure of $\sigma_{zz}$ and $\sigma_{zr}$ proposed by Franklin and Alvarez \cite{Franklin_7}. The first one is to take into account the redirection of forces through a constant coefficient (dimensionless) \cite{Duran} $\kappa$: $\sigma_{zr}=\mu_s\kappa\sigma_{zz}$, where $\mu_s\approx\tan(32^o)$ is the friction coefficient between the grains, and the grains and the pipe walls. The second is to model $\sigma_{zr}$ as a function of the square of the grains velocity: $\sigma_{zr} = 1/2 \rho_s\mu_s {v_s}^2$. The third is to consider that capillary forces can be modeled as a multiplicative constant $b \geq 1$ on the friction term. This is a simple way to take into account capillary forces, that act in the same direction as the friction forces. However, for this study, we fixed $b=1$ and did not change it. The resulting equation is

\begin{equation}
\frac{\partial cv_{s}}{\partial t} + \frac{\partial c v_{s}^{2}}{\partial z}  = c\, g\, \cos\,\theta - \frac{1}{\rho_{s}} \frac{\partial P}{\partial z}  - \frac{v_{s}}{\kappa} \frac{\partial v_{s}}{\partial z} - \frac{2}{D}\mu_s v_{s}^2
\label{eq:motion}
\end{equation}

Normalizing Eq. \ref{eq:motion} by the characteristic length $L_c$, time $t_c$, velocity $v_c$ and pressure $P_c = \rho_sgD$, we obtain Eq. \ref{eq:motion2}

\begin{equation}
c \frac{\partial v_{s}^*}{\partial t^*} +v_{s}^* \left( \frac{\partial c}{\partial t^*} + v_{s}^* \frac{\partial c}{\partial z^*} + c \frac{\partial v_{s}^*}{\partial z^*} \right)  + c v_{s}^* \frac{\partial v_{s}^*}{\partial z^*} = c\, \cos \,\theta - \frac{\partial P^*}{\partial z^*}  - \frac{v_{s}^*}{\kappa} \frac{\partial v_{s}^*}{\partial z^*} - 2  \mu_s (v_{s}^*)^2
\label{eq:motion2}
\end{equation}

In Eq. \ref{eq:motion2} we identify the normalized mass conservation equation (Eq. \ref{eq:mass2}); therefore, with an additional simplification, we obtain Eq. \ref{eq:motion3}.

\begin{equation}
c \frac{\partial v_{s}^*}{\partial t^*} + c v_{s}^* \frac{\partial v_{s}^*}{\partial z^*} - c\, \cos \,\theta + \frac{\partial P^*}{\partial z^*} + \frac{v_{s}^*}{\kappa} \frac{\partial v_{s}^*}{\partial z^*} + 2  \mu_s (v_{s}^*)^2 = 0
\label{eq:motion3}
\end{equation} 

\noindent where $P^*=P/(\rho_sgD)$ 

\subsection{Air pressure}

For the air pressure, an equation based on the work of Bertho et al. \cite{Bertho_2} is used. Bertho et al. \cite{Bertho_2} combined the mass conservation equations for the air and grains, the isentropic relation for the air, and Darcy's equation relating the air flow through packed grains to the pressure gradient to obtain Eq. \ref{eq:pressure}

\begin{equation}
\frac{\partial P}{\partial t} + v_s\frac{\partial P}{\partial z} + \frac{\gamma P}{(1-c)}\frac{\partial v_{s}}{\partial z} - B\frac{\partial^2 P}{\partial z^2}=0
\label{eq:pressure}
\end{equation}

\noindent where $\gamma$ is the ratio of specific heats ($1.4$ for air) and $B$ is a coefficient given by

\begin{equation}
B = \frac{\gamma P\left( 1-c\right)^2d^2}{\mu_a180c^2}
\label{eq:pressure_2}
\end{equation}

\noindent where $\mu_a$ is the dynamic viscosity of air. In Eq. \ref{eq:pressure_2}, $B$ was obtained by estimating the permeability of grains using the Carman--Kozeny equation. Normalizing Eq. \ref{eq:pressure} by the characteristic length $L_c$, time $t_c$, velocity $v_c$ and pressure $P_c$, we obtain Eq. \ref{eq:pressure3}

\begin{equation}
\frac{\partial P^*}{\partial t^*} = - v_{s}^*\frac{\partial P^*}{\partial z^*} - \frac{\gamma P^*}{(1-c)}\frac{\partial v_{s}^*}{\partial z^*} + B^*\frac{\partial^2 P^*}{\partial z^{*2}}
\label{eq:pressure3}
\end{equation}

\noindent where $B^*=Bg^{-1/2}D^{-3/2}$ is a dimensionless coefficient.

\section{Stability analysis}
\label{section:analysis}

A linear stability analysis is presented based on Eqs. \ref{eq:mass2}, \ref{eq:motion3} and \ref{eq:pressure3}, which are solved for $P^*$, $c$ and $v^*$. The main objective is to find the typical length for the high-density regions of the granular flow. The initial state, considered as the basic state, is a steady, dense uniform flow of grains known as compact regime \cite{Aider}. This state is then perturbed and we investigate if a preferential mode exists, i.e., we investigate if an initial compact regime will be fractured in granular plugs with a preferential wavelength. Thus, the analysis considers a basic state in which the pressure is equal to the characteristic pressure, $P_c$, the grain velocity is equal to the characteristic velocity $v_c$, and the compactness is equal to an average dense compactness $c_0$ (where, in compact regime $c_0 \approx 0.55$). The pressure, grain velocity and compactness are then the sum of the basic state, of $O(1)$, and the perturbation, of $O(\epsilon)$, $\epsilon\ll 1$. In dimensionless form:

\begin{equation}
P^* = 1 + \tilde{P}, 
\qquad
v_{s}^* = 1 + \tilde{v_{s}},
\qquad
c = c_0 + \tilde{c}
\label{eq:basic_pert}
\end{equation}

\noindent where $\tilde{P}\,\ll\,1$, $\tilde{v}\,\ll\,1$ and $\tilde{c}\,\ll\,1$ are respectively the pressure, velocity and compactness perturbations (dimensionless).

By inserting the pressure, the velocity and the compactness from Eq. \ref{eq:basic_pert} in Eqs. \ref{eq:mass2}, \ref{eq:motion3} and \ref{eq:pressure3}, and keeping only the terms of $O(\epsilon)$, we obtain

\begin{equation}
\frac{\partial \tilde{c}}{\partial t^*} + c_0 \frac{\partial \tilde{v_{s}}}{\partial z^*} + \frac{\partial \tilde{c}}{\partial z^*} = 0
\label{eq:pert_mass}
\end{equation}

\begin{equation}
c_0 \frac{\partial \tilde{v_{s}}}{\partial t^*} + c_0 \frac{\partial \tilde{v_{s}}}{\partial z^*} -\tilde{c}\,\cos\,\theta + \frac{\partial \tilde{P}}{\partial z^*} + \frac{1}{\kappa} \frac{\partial \tilde{v_{s}}}{\partial z^*} + 4\mu_s \tilde{v_{s}} = 0
\label{eq:pert_motion}
\end{equation}

\begin{equation}
\frac{\partial \tilde{P}}{\partial t^*} = - \frac{\partial \tilde{P}}{\partial z^*} - \frac{\gamma}{(1-c_0)} \frac{\partial \tilde{v_{s}}}{\partial z^*} + B_{1}^* \frac{\partial^2 \tilde{P}}{\partial z^{*2}}
\label{eq:pert_pressure}
\end{equation}

\noindent where $B_1^*$ is a constant obtained by replacing $P$ by $P_c$ in $B^*$. Equations \ref{eq:pert_mass}, \ref{eq:pert_motion} and \ref{eq:pert_pressure} form a linear system with constant coefficients; therefore, the solutions can be found by considering the following normal modes:

\begin{equation}
\begin{array}{c} \tilde{c} = \hat{c}\, e^{i(k^*z^* - \omega^*t^*)} + c.c.\\
\\ \tilde{v_{s}} = \hat{v_{s}}\, e^{i(k^*z^* - \omega^*t^*)} + c.c.\\
\\ \tilde{P} = \hat{P}\, e^{i(k^*z^* - \omega^*t^*)} + c.c.\\ \end{array}
\label{normal_modes}
\end{equation}

\noindent where $k^*=kD=2\pi D/\lambda \in \mathbb{R}$ is the dimensionless wavenumber in the $z^*$ direction, $\lambda$ is the wavelength in the $z^*$ direction, $\hat{c}\in \mathbb{C}$ $\hat{v}\in \mathbb{C}$ and $\hat{P}\in \mathbb{C}$ are the dimensionless amplitudes, and $c.c.$ stands for the complex conjugate. Let $\omega^*\in \mathbb{C}$, $\omega^*\,=\,\omega_r^*\,+\,i\omega_i^*$, where $\omega_r^* = \omega_r/(kv_c)\in \mathbb{R}$ is the dimensionless angular frequency and $\omega_i^* = \omega_it_c \in \mathbb{R}$ is the dimensionless growth rate. By inserting the normal modes in Eqs. \ref{eq:pert_mass}, \ref{eq:pert_motion} and \ref{eq:pert_pressure}, we obtain Eq. \ref{eq:eigenvalue}.

\begin{equation}
\begin{array}{c} \left[\begin{array}{ccc} ik^*-i\omega^* & ik^* c_0 & 0 \\ -\cos\,\theta & -i\omega^* c_0 + ik^* c_0 +\kappa^{-1} ik^* + 4\mu_{s}  & ik^*\\  0 &  \frac{\gamma} {1-c_0} ik^* & -i\omega^* + ik^* +B_1^* k^{*2} \\ \end{array}\right]  \left[\begin{array}{c} \hat{c} \\ \hat{v_{s}} \\ \hat{P} \\ \end{array} \right] \,= \\ \, \\= \ \left[\begin{array}{c} 0 \\ 0 \\ 0 \\ \end{array} \right]\end{array}
\label{eq:eigenvalue}
\end{equation}

The existence of non-trivial solutions for this system requires its determinant to be zero. This results in

\begin{equation}
\begin{array} {c} i c_0\,\omega^{*3} + \omega^{*2}\left[k^*(-3ic_0-i\kappa^{-1}) - k^{*2} c_0\,B_1^* - 4\mu_s\right] + \\ \, \\ +\omega^* [ k^{*3}(2 c_0 \,B_1^* + \kappa^{-1} B_1^*) + ik^{*2} (2 \kappa^{-1} - \frac{\gamma}{1-c_0} +3 c_0-4\mu_{s} B_1^*) \\ \, \\ + k^*(c_0\, \cos\,\theta\,+8\mu_{s}) ] + \\ \, \\ +ik^{*3}\left(-c_0 - \kappa^{-1} + 4\mu_{s} B_1^* + c_0\, \cos\,\theta\, B_1^{*} + \frac{\gamma}{1-c_0}\right) + \\ \, \\ + k^{*2} \left( -4\mu_s - c_0\, \cos \, \theta\right) + B_1^* k^{*4} \left( -\kappa^{-1} - c_0\right)   = 0 \\ \end{array}
\label{eq:det}
\end{equation}

Equation \ref{eq:det} is solved to find $\omega^*(k^*)$. In order to solve it, constant $b$ was assumed to be equal to $1$. Constant $B_1^*$ was computed using the characteristic values of our experiments, i.e., $D=3\,mm$, $d = d_{50} = 0.225\,mm$ (the latter value obtained by SEM), and we assumed that $\mu_s = \tan (32^o)$, $\kappa \approx 0.5$, and $c_0\approx 0.55$ \cite{Duran,Cambau}. Additionally, we assumed $\theta \leq 5^o$. The imaginary part of $\omega^*(k^*)$, $\omega_i^*(k^*)$, was investigated in order to obtain the typical length of granular plugs.

\begin{figure}
\begin{minipage}[c]{\textwidth}
\centering
\begin{tabular}{c}
\includegraphics[scale=0.65]{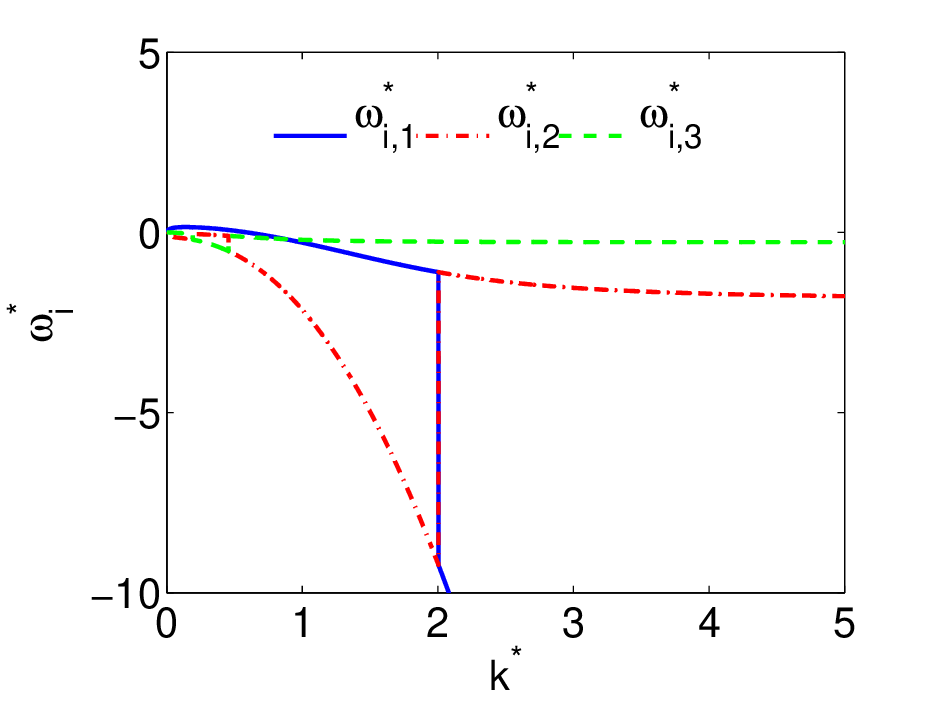}\\
      (a)
\end{tabular}
\end{minipage} \hfill
\begin{minipage}[c]{\textwidth}
\centering
\begin{tabular}{c}
\includegraphics[scale=0.65]{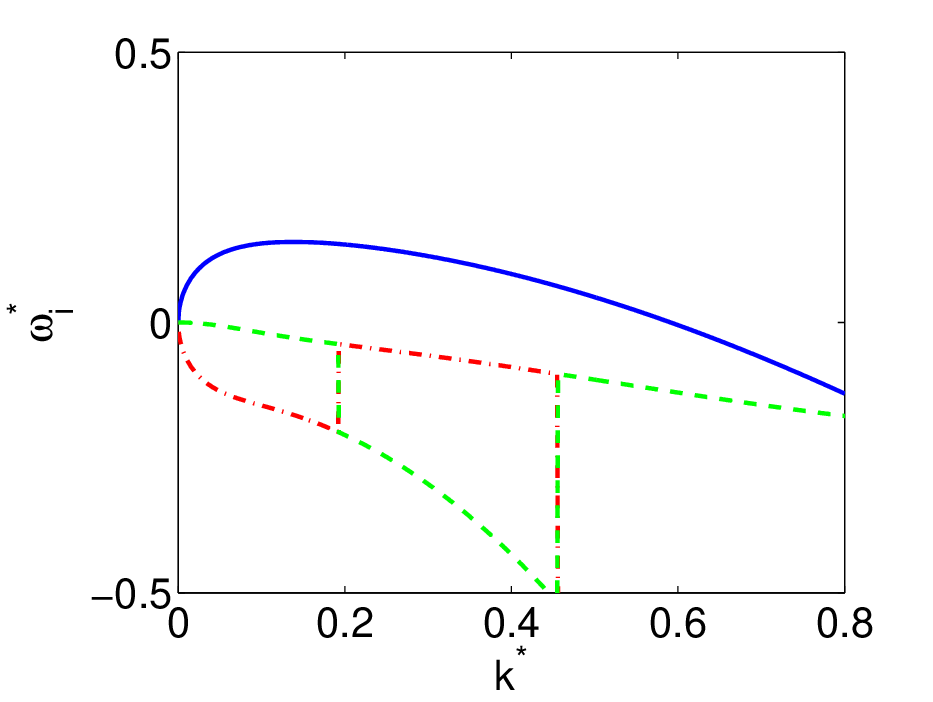}\\
      (b)
\end{tabular}
\end{minipage}
\caption{Dimensionless growth rate $\omega_i^*$ as a function of dimensionless wavenumber $k^*$. The continuous line corresponds to one root of Eq. \ref{eq:det}, the dashed-dot line and the dashed line to the others.}
\label{fig:result_stability}
\end{figure}

Figure \ref{fig:result_stability} shows the dimensionless growth rate $\omega_i^*$ as a function of dimensionless wavenumber $k^*$. The continuous line corresponds to one root of Eq. \ref{eq:det}, the dashed-dot line and the dashed line to the others. Figure \ref{fig:result_stability}a, for the broad range of wavenumbers, shows that small wavelengths are stable. Figure \ref{fig:result_stability}b illustrates the $0\lesssim kD\lesssim 0.8$ region. The figure shows there is a solution, given by the continuous line, which corresponds to a long-wavelength instability, with a preferential mode in $k^*\,\approx\,0.2$ and a cut-off wavelength of $k^*\,\approx\,0.6$. This corresponds to wavelengths in the order of $10D$.

\section{Discussion}
\label{section:Discussion}

These experiments and those of Franklin and Alvarez \cite{Franklin_7} showed that the plug sizes are $3\, <\,\lambda /D\, <\, 11$, which is in perfect agreement with the proposed model. However, as only one tube diameter was employed, we then compare these results with the previously published results.

In a series of papers, Raafat et al. \cite{Raafat}, Aider et al. \cite{Aider}, and Bertho et al. \cite{Bertho_1} presented experiments of granular flows through a tube. In particular, with regard to the characteristics of density waves, Raafat et al. \cite{Raafat} reported the size of plugs was $\lambda /D \approx 10$ and that it was approximately independent of the flow rate. Bertho et al. \cite{Bertho_1} also reported that the size of plugs was $\lambda /D \approx 10$. In addition, they showed the length of air bubbles is $\lambda_{bubble} /D \approx 10$. These measurements are in agreement with the lengths predicted by the proposed model.

Because the dense waves appearing in the vertical chute of grains in narrow tubes share similarities with the waves appearing in the vertical pneumatic conveying in dense regime, comparisons with some experiments on pneumatic conveying are presented next. Borzone and Kinzing \cite {Borzone} reported plugs with $2\, <\,\lambda /D\, <\, 15$ while Jaworski and Dyakowski \cite{Jaworski} reported the existence of series of granular plugs, where each plug has a length between $2D$ and $4D$, in agreement with the proposed model.

The results of the model are in agreement with the results reported by Franklin and Alvarez \cite{Franklin_7}, even though in the present study  the compactness $c$ was assumed as a variable, different from that work, where the authors fixed $c$ as constant. In addition, through $\theta$ it is possible to consider small deviations of the tube with respect to the vertical alignment. This allows to take into consideration small angle variations that may have occurred in the cited works.

Peng and Herrmann \cite{Peng}, using lattice-gas automata, studied density waves in vertical pipes. The authors reported that both the energy dissipation and the roughness of the walls of the pipe are essential to the formation of density waves. However, they did not measure the typical length of granular plugs. Poschell \cite{Poschel} investigated the formation of density waves using 2D Molecular Dynamics simulations.  The author found that the kinetic energy of the falling grains increases up to a characteristic threshold, from which density waves of no definite wavelength appear. Ellingsen et al. \cite{Ellingsen} studied the gravitational flow of grains through a narrow pipe under vacuum conditions. Their numerical results showed that granular waves could form in the absence of air if the dissipation caused by the collisions among the grains was smaller than that between the grains and the walls. However, the proposed model cannot predict the wavelength of the density waves in the presence of interstitial gas. Lately, Verb{\"u}cheln et al. \cite{Verbucheln}, using particle-based numerical simulations, studied a method to homogenize granular flows in order to avoid the formation of density waves. They did not report the value of the ratio between the plug length and the tube's diameter, $\lambda /D$; however, in their simulations the granular plugs usually appeared when the diameter was $3$ mm.

The present model is based on the work of Bertho et al. \cite{Bertho_2}, with the inclusion of closure equations for the friction terms. Bertho et al. \cite{Bertho_2} solved numerically their model equations and found that the formation of a decompaction wave was captured by the model; however, the numerical simulations did not explain the length of plugs that may appear in narrow pipes. Different from Bertho et al. \cite{Bertho_2}, we included closure equations that allowed us to perform a linear stability analysis, which showed the existence of a most unstable mode. This analysis explains the typical length of plugs appearing in narrow pipes.

As far as we know, Lee \cite{Lee} performed the only stability analysis previous to our work. In his analysis, Lee neglected air effects (pressure and drag); therefore, the analysis was not able to find the correct length scale of plugs. Different from Lee \cite{Lee}, we considered air effects, and the present stability analysis predicts a wavelength that agrees with typical lengths observed in different experiments.

The present analysis considers only one median diameter as the scale for the grain size. Although dispersion around the median diameter is admitted, it is expected that the analysis may fail in the case of large dispersions around the medium value. If this is the case, other scales for the grain size may be necessary. In addition, cohesion among grains may be present. In this problem, cohesion has two sources: (i) capillarity, when air relative humidity is high, commonly above $70\%$, and (ii) electrostatic forces, when air relative humidity is low, commonly below $30\%$. While the model considers capillarity via a multiplicative factor (Subsection \ref{section_granular_motion}), it does not consider electrostatic forces. The present paper is concerned with air relative humidity between $30\%$ and $70\%$, so that these cohesion effects are not very strong; however, we understand that further investigations on cohesion forces, as well as on polydisperse grains, are necessary in order to fully understand the problem.

We note that this study is interested in the formation and dynamics of granular plugs. However, the formation of plugs must be avoided in many industrial applications. In these cases, the formation of plugs can be prevented by controlling the air relative humidity or the flow rate of gains.

The final observation concerns the lowest plug in the gravitational dense flow. Bertho et al. \cite{Bertho_1} reported that at the lower portion of the tube (tube exit) a different plug is formed. The length of this plug varies with the flow rate. For $\dot{m}$ from $1.75\,g/s$ to $3.9\,g/s$, they found that the length of the bottom plug varies from $\lambda /D \approx 30$ to $\lambda /D \approx 200$. This plug is subject to exit boundary conditions; therefore, its length is not correctly predicted by the present analysis.

\section{Conclusions}
\label{section:conclusions}

This paper focused on the density waves that appear when fine grains fall through a narrow tube. Its main objective was to experimentally determine the wavelengths and celerities of density waves that appear when fine grains fall through vertical and slightly inclined pipes. In our experiments, different sized glass spheres flowed driven by gravity through a vertical glass pipe, and the granular flow was filmed with a high-speed camera. The ambient temperature and relative humidity were controlled. Both the mean celerity and wavelengths were calculated using an image processing code developed in MatLab environment. When varying the mass flow rate, two wave regimes occurred: the density waves either propagated at a constant celerity or oscillated over a mean drift celerity. The oscillations suggest a stick-slip displacement of the plugs. The density waves appeared with mass flow rates between 0.1 and 0.95 $g/s$, ten times smaller than previously reported values. The paper also presented a stability analysis based on equations proposed by Bertho et al. \cite{Bertho_2}, with small modifications and in dimensionless form. In our analysis, the basic state is a compact regime, and we investigated if it would be fractured in granular plugs with a preferential wavelength. The stability analysis predicts a wavelength in the order of $10D$ for the high-density regions. This predicted length scale is in good agreement with our experimental results as well as with previously published results.

\section{Acknowledgments}

\begin{sloppypar}
Carlos A. Alvarez is grateful to the Ecuadorian government foundation SENESCYT for the scholarship grant (no. 2013-AR2Q2850). Erick de Moraes Franklin is grateful to FAPESP (grant no. 2012/19562-6), to CNPq (grant no. 471391/2013-1) and to FAEPEX/UNICAMP (conv. 519.292, project AP0008/2013) for the financial support provided. 
\end{sloppypar}






\bibliographystyle{elsarticle-num}
\bibliography{references}

\begin{thebibliography}{10}
\expandafter\ifx\csname url\endcsname\relax
  \def\url#1{\texttt{#1}}\fi
\expandafter\ifx\csname urlprefix\endcsname\relax\def\urlprefix{URL }\fi
\expandafter\ifx\csname href\endcsname\relax
  \def\href#1#2{#2} \def\path#1{#1}\fi

\bibitem{Duran}
J.~Duran, Sands, powders and grains: an introduction to the physics of granular
  materials, 2nd Edition, Springer, 1999.

\bibitem{Campbell}
C.~Campbell, Granular material flows \mbox{-} an overview, Powder Technology
  162~(3) (2006) 208 -- 229.

\bibitem{Elbelrhiti}
H.~Elbelrhiti, P.~Claudin, B.~Andreotti, Field evidence for
  surface-wave-induced instability of sand dunes, Nature 437~(04058).

\bibitem{Franklin_6}
E.~M. Franklin, Linear and nonlinear instabilities of a granular bed:
  determination of the scales of ripples and dunes in rivers, Appl. Math.
  Model. 36 (2012) 1057--1067.

\bibitem{GDR_midi}
GDR-MiDi, On dense granular flows, The European Physical Journal E 14~(4).
\newblock \href {http://dx.doi.org/10.1140/epje/i2003-10153-0}
  {\path{doi:10.1140/epje/i2003-10153-0}}.

\bibitem{Jaeger}
H.~M. Jaeger, S.~R. Nagel, Physics of the granular state, Science 255~(5051)
  (1992) 1523--1531.

\bibitem{Aider}
J.-L. Aider, N.~Sommier, T.~Raafat, J.-P. Hulin, Experimental study of a
  granular flow in a vertical pipe: A spatiotemporal analysis, Phys. Rev. E 59
  (1999) 778--786.

\bibitem{Bertho_1}
Y.~Bertho, F.~Giorgiutti-Dauphin{\'e}, T.~Raafat, E.~J. Hinch, H.~J. Herrmann,
  J.~P. Hulin, Powder flow down a vertical pipe: the effect of air flow, J.
  Fluid Mech. 459 (2002) 317--345.

\bibitem{Raafat}
T.~Raafat, J.~P. Hulin, H.~J. Herrmann, Density waves in dry granular media
  falling through a vertical pipe, Phys. Rev. E 53 (1996) 4345--4350.

\bibitem{Savage}
S.~B. Savage, Gravity flow of cohesionless granular materials in chutes and
  channels, J. Fluid Mech. 92 (1979) 53--96.

\bibitem{Wang}
C.~H. Wang, R.~Jackson, S.~Sundaresan, Instabilities of fully developed rapid
  flow of granular material in channel, J. Fluid Mech. 342 (1997) 179--197.

\bibitem{Lee}
J.~Lee, Density waves in the flows of granular media, Phys. Rev. E 49 (1994)
  281--298.

\bibitem{Bagnold_4}
R.~A. Bagnold, Experiments on a gravity-free dispersion of large solid spheres
  in a newtonian fluid under shear, Proc. R. Soc. London Ser. A 225~(1160)
  (1954) 49--63.

\bibitem{Franklin_7}
E.~M. Franklin, C.~A. Zambrano, Length scale of density waves in the
  gravitational flow of fine grains in pipes, J. Braz. Soc. Mech. Sci. Eng.
  37~(5) (2015) 1507--1513.
\newblock \href {http://dx.doi.org/10.1007/s40430-014-0291-3}
  {\path{doi:10.1007/s40430-014-0291-3}}.

\bibitem{Ellingsen}
S.~A. Ellingsen, K.~S. Gjerden, M.~Gr\o{}va, A.~Hansen, Model for density waves
  in gravity-driven granular flow in narrow pipes, Phys. Rev. E 81 (2010)
  061302.

\bibitem{Verbucheln}
F.~Verb{\"u}cheln, E.~J. Parteli, T.~P{\"o}schel, Helical inner-wall texture
  prevents jamming in granular pipe flows, Soft matter 11~(21) (2015)
  4295--4305.

\bibitem{Konrad_1}
K.~Konrad, Dense-phase pneumatic conveying: {A} review, Powder Technology 49
  (1986) 1--35.

\bibitem{Borzone}
L.~A. Borzone, G.~Klinzing, Dense-phase transport: vertical plug flow, Powder
  Technology 53 (1987) 273--283.

\bibitem{Konrad_2}
K.~Konrad, T.~S. Totah, Vertical pneumatic conveying of particle plugs, Can. J.
  Chem. Eng 67 (1989) 245--252.

\bibitem{Jaworski}
A.~J. Jaworski, T.~Dyakowski, Investigations of flow instabilities within the
  dense pneumatic conveying system, Powder Technology 125 (2002) 279--292.

\bibitem{Bertho_2}
Y.~Bertho, F.~Giorgiutti-Dauphin{\'e}, J.~P. Hulin, Intermittent dry granular
  flow in a vertical pipe, Physics of Fluids 15~(11) (2003) 3358--3369.

\bibitem{Poschel}
T.~P{\"o}schel, Recurrent clogging and density waves in granular material
  flowing through a narrow pipe, Journal de Physique I 4~(4) (1994) 499--506.

\bibitem{Janssen}
H.~Janssen, Versuche {\"u}ber getreidedruck in silozellen, Zeitschr. d.
  Vereines deutscher Ingenieure 39~(35) (1895) 1045--1049.

\bibitem{Cambau}
T.~Cambau, J.~Hure, J.~Marthelot, Local stresses in the janssen granular
  column, Phys. Rev. E 88 (2013) 022204.

\bibitem{Peng}
G.~Peng, H.~J. Herrmann, Density waves of granular flow in a pipe using
  lattice-gas automata, Phys. Rev. E 49~(3) (1994) R1796.

\end{thebibliography}







\end{document}